\documentclass[twocolumn, showpacs, preprintnumbers,
nofootinbib, aps, prd, superscriptaddress,
10pt, showkeys, notitlepage, longbibliography]{revtex4-1}

\usepackage[dvips]{graphicx}
\usepackage{bm,latexsym,amsmath,amssymb,amsfonts,times}
\usepackage{color}
\usepackage{natbib}
\usepackage{times}
\usepackage{graphicx}
\usepackage{dcolumn}
\usepackage{mathtools}

\def\be {\begin{equation}}
\def\ee {\end{equation}}
\def\ba {\begin{eqnarray}}
\def\ea {\end{eqnarray}}

\def\c  {\gamma}

\def\bi {\begin{itemize}}
\def\ei {\end{itemize}}

\newcommand\beq{\begin{eqnarray}}
\newcommand\eeq{\end{eqnarray}}

\usepackage[linktocpage,breaklinks]{hyperref}
\definecolor{venetianred}{rgb}{0.78, 0.03, 0.08}
\definecolor{grey}{rgb}{0.25, 0.25, 0.28}
\definecolor{darkmidnightblue}{rgb}{0.0, 0.2, 0.4}
\definecolor{egyptianblue}{rgb}{0.06, 0.2, 0.65}
\definecolor{darkblue}{rgb}{0.0, 0.0, 0.55}
\hypersetup{colorlinks,breaklinks,
            linkcolor=red,urlcolor=darkblue,
            anchorcolor=red,citecolor=blue}

\def\X5sp{{\rm X}_5}
\def\Y3sp{{\rm Y}_3}
\def\Z3sp{{\rm Z}_3}

\begin{document}

\title{Solutions in the generalized Proca theory with the nonminimal coupling to the Einstein tensor}

\author{Masato Minamitsuji}
\email{masato.minamitsuji@ist.utl.pt}
\affiliation{Centro Multidisciplinar de Astrofisica - CENTRA,
Instituto Superior Tecnico - IST,
Universidade de Lisboa - UL,
Avenida Rovisco Pais 1, 1049-001, Portugal}

\begin{abstract}
We investigate the static and spherically symmetric solutions in a class of the generalized Proca theory with the nonminimal coupling to the Einstein tensor. First, we show that the solutions in the scalar-tensor theory with the nonminimal derivative coupling to the Einstein tensor can be those in the generalized Proca theory with the vanishing field strength. We then show that when the field strength takes the nonzero value the static and spherically symmetric solutions can be found only for the specific value of the nonminimal coupling constant. Second, we investigate the first-order slow-rotation corrections to the static and spherically symmetric background. We find that for the background with the vanishing electric field strength the slowly rotating solution is identical to the Kerr- (anti-) de Sitter solutions in general relativity. On the other hand, for the background with the nonvanishing electric field strength the stealth property can realized at the first order in the slow-rotation approximation.
\end{abstract}
\date{\today}
\maketitle

%\tableofcontents
%%%%%%%%%%%%%%%%%%%%%%%%%
\section{Introduction}
\label{sec:intro}

Cosmological observations suggest the existence of the mysterious elements
in the history of the Universe,
such as the inflationary evolution at the beginning of the Universe,
and dark matter and dark energy at the present day.
General relativity (GR) 
with these new elements in the right-hand side of the Einstein equation
as the energy-momentum sources
are mathematically equivalent to the certain modification of GR 
where the left-hand side of the Einstein equation is modified
by the new gravitational degrees of freedom 
in addition to the metric tensor  \cite{Clifton:2011jh,Berti:2015itd}.
In many cases,
the modification of GR can be described by
a scalar-tensor theory of gravitation
at least in a certain regime~\cite{Fujii:2003pa}. 
Realistic modification of GR should not contain
the so-called Ostrogradsky ghosts 
associated with the higher-derivative interactions~\cite{Woodard:2015zca},
and
should be endowed with
the mechanisms that could suppress the extra gravitational degrees of freedom
around the locally gravitating sources~\cite{Vainshtein:1972sx,Brax:2004qh},
in order to be compatible with the tests of GR in the weak gravity regime.
Scalar-tensor theories that could satisfy all these requirements 
typically belong to the so-called generalized Galileon / Horndeski
theory~\cite{Horndeski:1974wa,Deffayet:2009mn,Deffayet:2011gz,Kobayashi:2011nu}
where the equations of motion are given by the second-order differential equations 
despite the existence of the higher derivative interactions in the Lagrangian.

While the scalar-tensor Horndeski theory has been extensively investigated,
it is also interesting to look for the similar theories for the other field species.
In this paper, we consider a class of the generalized vector-tensor theories of gravitation
where the equations of motion are given by the second-order differential equations.
It was shown that if the gauge symmetry of the vector field is preserved,
the Galileon-like extension of the vector field theory does not exist 
and only the Maxwell kinetic term is allowed \cite{Deffayet:2013tca}. 
A way out for this no-go theorem was to abandon the gauge symmetry.
The introduction of the mass term of the vector field breaks the gauge symmetry.
In the vector field theory with the mass term $m^2 A_\mu A^\mu$,
where $A_\mu$ is the vector field and the Greek indices $(\mu,\nu,...)$ run the four-dimensional spacetime, 
the so-called Proca theory,  
the vector field contains the three propagating degrees of freedom,
namely one longitudinal and two transverse degrees of freedom.
The generalization of the massive vector field theory to the Galileon-like theory
was first investigated in Refs. \cite{Tasinato:2014eka,Heisenberg:2014rta},
and then extended in Refs. \cite{Allys:2015sht,Jimenez:2016isa,DeFelice:2016cri,DeFelice:2016yws,DeFelice:2016uil}
including the generalization of the interaction of the field strength with the double dual of the Riemann tensor Ref. \cite{Horndeski1976}.
In the generalized Proca theory,
the screening mechanism and cosmology have been investigated in Refs. \cite{DeFelice:2016cri,DeFelice:2016yws,DeFelice:2016uil}.

In this paper, 
we will investigate the static and spherically symmetric solutions  
in the subclass of the generalized Proca theory
which possesses the nonminimal coupling of the vector field to the Einstein tensor
$G^{\mu\nu} A_\mu A_\nu$,
where $G_{\mu\nu}$ is the Einstein tensor associated with the metric $g_{\mu\nu}$.
First, we will show that 
the solutions in the scalar-tensor Horndeski theory with the nonminimal derivative coupling
to the Einstein tensor $G^{\mu\nu}\partial_\mu\varphi\partial_\nu\varphi$
\cite{Sushkov:2009hk,Saridakis:2010mf,Germani:2010gm,Germani:2011ua,Gubitosi:2011sg}
can also be those in the above generalized Proca theory
with the vanishing field strength.
In this subclass of the scalar-tensor Horndeski theory,
the static and spherically symmetric solutions have been obtained
in Refs. \cite{Babichev:2013cya,Rinaldi:2012vy,Minamitsuji:2013ura,Anabalon:2013oea}
(see also 
Refs. \cite{Kobayashi:2014eva,Charmousis:2014zaa,Babichev:2015rva} for the more general theories
and 
Ref. \cite{Silva:2016smx,Babichev:2016rlq} for the reviews),
and the solutions particularly relevant for astrophysics or cosmology 
are the stealth Schwarzschild and the Schwarzschild- (anti-) de Sitter solutions
which were originally obtained in Ref. \cite{Babichev:2013cya}.

On the other hand, 
the nonexistence of the black hole solutions with the massive vector field charge 
has been proven by Bekenstein in Refs. \cite{Bekenstein:1971hc,Bekenstein:1972ky}.
As shown in recent work \cite{Chagoya:2016aar}, however,
the no-hair argument can be avoided
once the nonminimal coupling $G^{\mu\nu} A_\mu A_\nu$ is introduced.
\footnote{
As argued in Ref.~\cite{Herdeiro:2016tmi},
the no-hair argument for the Proca theory can also be circumvented 
for the complex massive Proca field.}
This corresponds to the simplest class of the ghost-free bilinear nonminimal couplings of the vector field
to the divergence-free Lovelock tensors \cite{Geng:2015kvs}. 
%%%%%%%%%%%%%%%%%%
Moreover,
as argued in Ref.  \cite{Jimenez:2014rna},
the nonminimal coupling of the vector field to the Einstein tensor, $G^{\mu\nu} A_\mu A_\nu$,
can also arise from the quadratic gravitational theory in the Weyl geometry.
%%%%%%%%%%%%%%%%%%
References \cite{Chagoya:2016aar,Geng:2015kvs} have obtained the black hole solutions 
for the particular value of the nonminimal coupling constant.
Reference \cite{Fan:2016jnz} has investigated the black hole solutions in the generalized Proca theory
with the nonminimal coupling $R A^\mu A_\mu$,
where $R$ is the Ricci scalar curvature.
In this paper,
we will extend these former attempts in Refs. \cite{Chagoya:2016aar,Geng:2015kvs},
clarify the relations among the solutions,
and also investigate the first-order slow-rotation corrections within the Hartle-Thorne approximation \cite{Hartle:1967he,Hartle:1968si}
along the line of Refs. \cite{Maselli:2015yva,Cisterna:2015uya}.

The paper is organized as follows:
In Sec. \ref{sec:action}, we will introduce the generalized Proca theory with the nonminimal coupling to the Einstein tensor
and derive the equations of motion.
In Sec. \ref{sec:solutions}, 
we will show how the solutions in the scalar-tensor theory 
can be described in the generalized Proca theory with the vanishing field strength.
In Sec. \ref{sec:solutions2},
we will obtain the solutions with the Coulomb potential in the temporal component of the vector field.
In Sec. \ref{sec:others},
we will explore the solutions with other forms of the vector field.
In Sec. \ref{sec:slow},
we will investigate the first-order slow-rotation corrections 
to the static and spherically symmetric solutions.
The last section, Sec. \ref{sec:conclusions}, is devoted to giving the concluding remarks.

%%%%%%%%%%%%%%%%%%%%%%%%%%%%%%
\section{The generalized Proca theory with the nonminimal coupling to the Einstein tensor}
\label{sec:action}

In this paper, we consider the generalized Proca theory given by  
\begin{align}
\label{eq:action}
S
&=
\int d^4x \sqrt{-g}
\left[
\frac{m_p^2}{2}
\left(R-2\Lambda\right)
-\frac{1}{4}F_{\mu\nu}F^{\mu\nu}
\right.
\nonumber\\
&\left.
-\left(m^2 g_{\mu\nu}-\beta G_{\mu\nu}\right)
 A^\mu A^\nu
\right],
\end{align}
where
$g_{\mu\nu}$ is the metric tensor, 
$R$ and $G_{\mu\nu}$ are the Ricci scalar and the Einstein tensor associated with $g_{\mu\nu}$,
$m_p$ and $\Lambda$ are the reduced Planck mass and the cosmological constant, respectively.
$A_\mu$ is the vector field,
$m$ and $\beta$ are the mass and the (dimensionless) nonminimal coupling constant of the vector field,
and 
$F_{\mu\nu}=\partial_\mu A_\nu-\partial_\nu A_\mu$ is the field strength.
After partially integrating and ignoring the boundary terms,
the action \eqref{eq:action} can be rewritten as 
\begin{align}
\label{eq:action2}
S
&=
\int d^4x \sqrt{-g}
\left[
\frac{m_p^2}{2}
\left(R-2\Lambda\right)
-\frac{1}{4}F_{\mu\nu}F^{\mu\nu}
-m^2 g_{\mu\nu} A^\mu A^\nu
\right.
\nonumber\\
&\left.
+\beta 
\left(
 (\nabla_\mu A^\mu)^2
-\nabla_\mu A_\nu \nabla^\nu A^\mu
-\frac{1}{2}A_\mu A^{\mu} R
\right) 
\right].
\end{align}
According to the formulation in Refs. \cite{Heisenberg:2014rta,DeFelice:2016cri,DeFelice:2016yws},
the action \eqref{eq:action2} corresponds to the case that
\begin{align}
G_2(X)=2m^2 X-\Lambda m_p^2,
\quad 
G_4(X)= \frac{m_p^2}{2}+\beta X,
\end{align}
with $c_2=0$ and $G_3(X)=G_5(X)=0$ where we have defined $X:=-\frac{1}{2} g^{\mu\nu}A_\mu A_\nu$.
The action \eqref{eq:action} is quadratic in $A_\mu$ and hence reflection-symmetric under $A_\mu\to -A_\mu$.
Thus if a set of the metric and the vector field $(g_{\mu\nu},A_\mu)$ is a solution of the theory \eqref{eq:action2},
the set $(g_{\mu\nu},-A_\mu)$ is also a solution of it.

Varying the action \eqref{eq:action} with respect to 
the metric $g_{\mu\nu}$ and the vector field $A_\mu$,
respectively,
the Einstein equation and the vector field equation of motion are obtained by 
\begin{widetext}
\begin{subequations}
\label{eoms}
\begin{align}
\label{eom_a}
m_p^2
\left(
G_{\mu\nu}+\Lambda g_{\mu\nu}
\right)
&=
\left(
F_{\mu\rho}F_{\nu}{}^{\rho}
-\frac{1}{4}g_{\mu\nu}F^{\rho\sigma}F_{\rho\sigma}
\right)
+2m^2 \left(A_\mu A_\nu-\frac{1}{2}g_{\mu\nu} A^\rho A_\rho\right)
\nonumber\\
&+\beta \left(
A^\rho A_\rho G_{\mu\nu}
+A_\mu A_\nu R
\right)
\nonumber\\
&-\beta g_{\mu\nu}
\left[
 \left(\nabla_\rho A^\rho\right)^2
-2\nabla_\rho A_\sigma \nabla^\rho A^\sigma 
+\nabla_\rho A_\sigma \nabla^\sigma A^\rho 
-2A_\rho\Box A^\rho
+2 A^\rho \nabla_\rho\nabla_\sigma A^\sigma
\right]
\nonumber\\
&-2\beta
\left[
 \nabla_\mu A_\rho \nabla_\nu A^\rho
-\nabla_\rho A^\rho \nabla_{(\mu}A_{\nu)}
-\nabla_\rho A_{(\mu} \nabla_{\nu)} A^\rho
+\nabla_\rho A_\mu \nabla^\rho A_\nu
+ A_\rho \nabla_{(\mu}\nabla_{\nu)} A^\rho
\right.
\nonumber\\
&\left.
-A^\rho \nabla_\rho \nabla_{(\mu} A_{\nu)}
+A_{(\mu} \Box A_{\nu)}
-2 A_{(\mu}\nabla_{\nu)}\nabla_\sigma A^\sigma
+A_{(\mu|} \nabla_{\rho}\nabla_{|\nu)} A^\rho
\right],
\\
\label{eom_b}
\nabla_\mu F^{\mu\nu}
&=2 
\left(m^2g^{\mu\nu}-\beta G^{\mu\nu}\right) A_\mu.
\end{align}
\end{subequations}
\end{widetext}
As expected the equations of motion \eqref{eoms} are given by the second-order differential equations.
Acting $\nabla_\nu$ on Eq. \eqref{eom_b} with $\nabla_\nu\nabla_\mu F^{\mu\nu}=0$,
we obtain 
\begin{align}
\label{constr}
\nabla_\nu
\left[
\left(m^2g^{\mu\nu}-\beta G^{\mu\nu}\right) A_\mu
\right]
=0,
\end{align}
which gives the constraint relation 
among the four components of the vector field $A_\mu$,
leaving the three physical degrees of freedom,
namely one longitudinal and two transverse degrees of freedom.

%%%%%%%%%%%%%%%%%%%%%%%%%%%%%%%%%%%%
\section{From the scalar-tensor theory to the generalized Proca theory}
\label{sec:solutions}

In this section, we show
how the static and spherically symmetric solutions in the scalar-tensor theory \eqref{eq:action3} 
are expressed in the generalized Proca theory \eqref{eq:action}
with the vanishing electric field strength.

\subsection{The correspondence}

We assume that the vector field $A_\mu$ can be decomposed
into the part given by the gradient of the scalar function $\varphi$ 
and the remaining vector field part $B_\mu$,
\begin{align}
\label{expansion}
A_\mu=\partial_\mu \varphi + B_\mu.
\end{align}
Since obviously
the scalar function does not contribute to the field strength
$F_{\mu\nu}=\partial_\mu A_\nu-\partial_\nu A_\mu =\partial_\mu B_\nu -\partial_\nu B_\mu=: F^{(B)}_{\mu\nu}$,
plugging Eq. \eqref{expansion} into the action \eqref{eq:action}, 
we obtain
\begin{align}
\label{eq:action_int}
S
&=
\int d^4x \sqrt{-g}
\left[
\frac{m_p^2}{2}
\left(R-2\Lambda\right)
-\frac{1}{4}F^{(B)}_{\mu\nu}F^{(B)\mu\nu}
\right.
\nonumber \\
&\left.
-\left(m^2 g_{\mu\nu}-\beta G_{\mu\nu}\right)
 \left(\nabla^\mu\varphi +B^{\mu} \right)
 \left(\nabla^\nu\varphi +B^{\nu} \right)
\right].
\end{align}
In the case of $B_\mu= 0$,
namely when the vector field is given by the gradient of the scalar function $A_\mu= \partial_\mu \varphi$,
the action \eqref{eq:action_int}
reduces to the shift-symmetric scalar-tensor theory
with the nonminimal derivative coupling to the Einstein tensor,
\begin{align}
\label{eq:action3}
S
&=
\int d^4x \sqrt{-g}
\left[
\frac{m_p^2}{2}
\left(R-2\Lambda\right)
\right.
\nonumber\\
&\left.
-\left(m^2 g_{\mu\nu}-\beta G_{\mu\nu}\right)
 \nabla^\mu\varphi 
 \nabla^\nu\varphi 
\right],
\end{align}
which involves the metric $g_{\mu\nu}$ and the scalar field $\varphi$ as the physical degrees of freedom
\cite{Sushkov:2009hk,Saridakis:2010mf,Germani:2010gm,Germani:2011ua,Gubitosi:2011sg}.

The static and spherically symmetric black hole solutions in the scalar-tensor theory \eqref{eq:action3}
have been investigated in 
Refs. \cite{Babichev:2013cya,Rinaldi:2012vy,Minamitsuji:2013ura,Anabalon:2013oea,Silva:2016smx,Babichev:2016rlq}
under the static and spherically symmetric metric ansatz 
\begin{align}
\label{metric_ansatz}
ds^2=-f(r)dt^2+\frac{dr^2}{h(r)}+r^2(d\theta^2+\sin^2\theta d\phi^2),
\end{align}
where $t$ and $r$ are the time and radial coordinates,
$\theta$ and $\phi$ are the polar and azimuthal coordinates of the two-sphere,
respectively,
and $f(r)$ and $h(r)$ are the functions of only the radial coordinate $r$.
In the static and spherically symmetric background \eqref{metric_ansatz},
the vector field has the nonvanishing $t$ and $r$ components
\begin{align}
\label{proca_ansatz}
A_\mu=\left(A_0(r), A_1(r),0,0\right),
\end{align}
where $A_0(r)$ and $A_1(r)$ are also only the functions of $r$.
Because of the reflection symmetry of the generalized Proca theory \eqref{eq:action}
and the absence of the cross terms of $A_0(r)$ and $A_1(r)$
(and their derivatives)
in the Einstein equation \eqref{eom_a} under the ansatz Eqs. \eqref{metric_ansatz} and \eqref{proca_ansatz}, 
if a set $\left(A_0(r), A_1(r)\right)=\left(c_0(r),c_1(r)\right)$ is a solution,
other sets $\left(A_0(r), A_1(r)\right)=\left(c_0(r), -c_1(r)\right), \left(-c_0(r),c_1(r)\right), \left(-c_0(r), -c_1(r)\right)$
are also solutions.
In each case, among them we will show the two independent branches $\left(A_0(r), A_1(r)\right)=\left(c_0(r), \pm c_1(r)\right)$.

We then derive the condition that the solution in the generalized Proca theory \eqref{eq:action}
is also the solution in the scalar-tensor theory \eqref{eq:action3}
within the ansatz Eqs. \eqref{metric_ansatz} and \eqref{proca_ansatz}.
Imposing the condition that $B_\mu=0$,
only the nontrivial component of the field strength $F_{rt}= A_0'(r)=0$,
which by integration gives  
\begin{align}
\label{p}
A_0(r)= P,
\end{align}
where $P$ is the constant.
Then from Eq. \eqref{expansion} with $B_\mu=0$, we identify
\begin{align}
\label{derives}
\partial_t \varphi=P,
\quad 
\partial_r \varphi=A_1(r).
\end{align}
Further integrating Eq. \eqref{derives},
the scalar function $\varphi$ is found to take the form of 
\begin{align} 
\label{scalar_ansatz}
\varphi(t,r)= P\, t +\psi(r),
\quad 
\psi(r):=\int dr A_1(r).
\end{align}
The scalar function of the form \eqref{scalar_ansatz} is 
exactly the same as in the black hole solutions found in the scalar-tensor theory \eqref{eq:action3}.
(See Ref. \cite{Babichev:2013cya} 
for $P\neq 0$
and Refs. \cite{Rinaldi:2012vy,Minamitsuji:2013ura,Anabalon:2013oea} 
for $P=0$.)
Thus Eq. \eqref{derives} gives how to express the solution 
in the scalar-tensor theory \eqref{eq:action3} in the generalized Proca theory \eqref{eq:action}
with the vanishing field strength.

On the other hand, 
the solution with the nonconstant $A_0(r)$,
giving rise to the nonvanishing electric field strength $F_{rt}=A_0'(r)$,
does not contain the counterpart in the scalar-tensor theory \eqref{eq:action3}.
In the rest of this section, 
we focus on the case that $B_\mu=0$ 
and show how the solutions in the scalar-tensor theory \eqref{eq:action3} 
discussed in Refs. \cite{Babichev:2013cya,Rinaldi:2012vy,Minamitsuji:2013ura,Anabalon:2013oea}
can be expressed in the generalized Proca theory \eqref{eq:action}.

\subsection{The stealth Schwarzschild solution}

The first example of the static and spherically symmetric solution
in the scalar-tensor theory \eqref{eq:action3}
is the stealth Schwarzschild solution obtained for $m=\Lambda=0$ \cite{Babichev:2013cya}.
In the generalized Proca theory \eqref{eq:action}, 
for general $P$ in Eq. \eqref{p} the solution is given by
\begin{subequations}
\label{eq:stealth}
\begin{align}
f(r)&
=h(r)
=1-\frac{2M}{r},
\\
A_1(r)&=
\pm \sqrt{\frac{2M}{r}}\frac{P}{f},
\end{align}
\end{subequations}
where $M$ is the integration constant that physically corresponds to the mass of the black hole.
The parameter $M$ appearing in the solutions discussed in the rest
also has the same physical meaning.

This is the stealth black hole solution 
in the sense that the amplitude of the vector field $P$ does not appear in the metric. 
Introducing the tortoise coordinate $dr_\ast=\frac{dr}{f}$,
\begin{align}
A_\mu dx^\mu
=P\left(dt\pm \sqrt{\frac{2M}{r}} dr_\ast\right),
\end{align}
which near the event horizons $r= 2M$ reduces to
\begin{align}
\label{stealth_vector}
A_\mu dx^\mu
\approx 
P\left(dt\pm dr_\ast\right)
=
P
\times 
  \begin{cases}
    dv \\
    du ,
  \end{cases}
\end{align}
where $v:=t+r_\ast$ and $u:= t-r_\ast$
are the advanced and retarded null coordinates.
The null coordinates $v$ and $u$ are regular at the future and past event horizons,
respectively,
ensuring the regularity of the scalar field there for each branch
in the context of the scalar-tensor theory \eqref{eq:action3}
\cite{Babichev:2013cya,Kobayashi:2014eva,Babichev:2015rva}.

\subsection{The Schwarzschild- (anti-) de Sitter  solution}
\label{sec:selftune}

Similarly, for $P=\pm \frac{m_p}{m} \sqrt{\frac{m^2+\beta\Lambda}{2\beta}}$ in Eq. \eqref{p} 
the Schwarzschild- (anti-) de Sitter solution
in the scalar-tensor theory \eqref{eq:action3} obtained in Ref. \cite{Babichev:2013cya}
is also expressed in the generalized Proca theory by
\begin{subequations}
\label{eq:selftuned}
\begin{align}
\label{eq:selftuned1}
f(r)&
=h(r)
=1-\frac{2M}{r}+\frac{m^2}{3\beta}r^2,
\\
A_1(r)&
=
\pm
\frac{m_p}{m}
\sqrt{-\frac{(m^2+\beta\Lambda) (m^2 r^3-6M\beta)}{6\beta^2 r}}
\frac{1}{f(r)},
\end{align}
\end{subequations} 
where the bare value of the cosmological constant $\Lambda$ does not appear 
in the metric functions $f(r)$ and $h(r)$,
and from the metric functions \eqref{eq:selftuned1} the effective cosmological constant is read as
$\Lambda_{\rm eff}=-\frac{m^2}{\beta}$.
Thus the spacetime is either asymptotically de Sitter or anti- de Sitter 
for $\beta<0$ and $\beta>0$,
respectively.
The positivity inside the square root in $A_0(r)$ requires
$\Lambda\geq -\frac{m^2}{\beta}$, irrespective of the sign of $\beta$.
Thus for $\beta>0$, $\Lambda$ can be either positive or negative,
while for $\beta<0$, $\Lambda$ is always positive.
For $m^2= -\beta\Lambda$,
$A_1(r)$ vanishes and 
the solution \eqref{eq:selftuned} reduces to the Schwarzschild- anti- de Sitter  in GR
with the cosmological constant $\Lambda$, 
\begin{align}
\label{sch_ds}
f(r)=h(r)=1-\frac{2M}{r}-\frac{\Lambda}{3}r^2.
\end{align}

Introducing the tortoise coordinate $dr_\ast=\frac{dr}{f}$,
\begin{align}
A_\mu dx^\mu
=
\frac{m_p}{m}
\sqrt{\frac{m^2+\beta\Lambda}{2\beta}}
\left(dt\pm \sqrt{\frac{-m^2 r^3+6M\beta}{3\beta r}} dr_\ast\right),
\end{align}
which near the (either event or cosmological) horizons $r\approx r_\ast$ 
satisfying $-m^2r_\ast^3+6M\beta=3\beta r_\ast$ 
becomes
\begin{align}
A_\mu dx^\mu
&\approx 
\frac{m_p}{m}
\sqrt{\frac{m^2+\beta\Lambda}{2\beta}}
\times 
  \begin{cases}
    dv \\
    du ,
  \end{cases}
\end{align}
where $v$ and $u$ are defined as in Eq. \eqref{stealth_vector}.
In the limit of $\beta\to-\frac{m^2}{\Lambda}$ the vector field trivially vanishes and 
the Schwarzschild- (anti-) de Sitter solution in GR with the cosmological constant $\Lambda$ is recovered.
The null coordinate $v$ is regular at the future event and past cosmological (only for $\beta<0$) horizons,
while 
the null coordinate $u$ is regular at the past event and future cosmological (only for $\beta<0$) horizons,
ensuring the regularity of the scalar field there for each branch
in the context of the scalar-tensor theory \eqref{eq:action3}
\cite{Babichev:2013cya,Kobayashi:2014eva,Babichev:2015rva}.

\subsection{The asymptotically anti- de Sitter  solution}

Finally,
for $P=0$ in Eq. \eqref{p} 
where in the theory Eq. \eqref{eq:action3} the scalar field $\varphi$ is time independent,
the asymptotically anti- de Sitter solution obtained in Refs. \cite{Rinaldi:2012vy,Minamitsuji:2013ura,Anabalon:2013oea}
in the theory \eqref{eq:action3}
is expressed in the generalized Proca theory by
\begin{subequations}
\label{nonflat}
\begin{align}
f(r)&=\frac{1}
              {3mr\beta \left(m^2-\beta\Lambda\right)^2}
\left[
m^7 r^3
-3m r\beta^3\Lambda^2
\right.
\nonumber\\
&\left.
+m^3 r\beta^2\Lambda \left(-6+r^2\Lambda\right) 
+m^5\beta \left(9r-2r^3\Lambda -24M\right)
\right.
\\
&\left.
+3\beta^{\frac{3}{2}}
\left(m^2+\beta\Lambda\right)^2{\rm arctan}\left(\frac{mr}{\sqrt{\beta}}\right)
\right],
\nonumber\\
h(r)&=\frac{ \left(m^2-\beta\Lambda\right)^2\left(m^2r^2+\beta\right)^2}
              {m^4\left(m^2r^2+\beta \left(2-r^2\Lambda\right)\right)^2}
         f(r),
\\
\label{adsa1}
A_1(r)
&=
\pm
\sqrt{
-\frac{m^2+\beta\Lambda}
        {2\beta \left(m^2r^2+\beta\right)h(r)}}
m_p r.
\end{align}
\end{subequations}
We require  $\beta>0$ so that the domain of $r$ is given by $0<r<\infty$.
Then, 
in order for $A_1(r)$ to be real outside the event horizon $h(r)>0$,
from Eq. \eqref{adsa1}
we find $\Lambda\leq -\frac{m^2}{\beta}$.

From the large-$r$ limit of the metric functions,
\begin{subequations}
\begin{align}
f(r)&\approx \frac{m^2r^2}{3\beta}
               +\frac{3m^2+\beta\Lambda}
                       {m^2-\beta\Lambda}
    +{\cal O}\left(\frac{1}{r}\right),
\\
h(r)&\approx \frac{m^2r^2}{3\beta}
               +\frac{7m^2+\beta\Lambda}
                       {3\left(m^2-\beta\Lambda\right)}
    +{\cal O}\left(\frac{1}{r}\right),
\end{align}
\end{subequations}
we find that 
the effective cosmological constant is
given by $\Lambda_{\rm eff}=-\frac{m^2}{\beta}<0$, 
and hence the spacetime is asymptotically anti- de Sitter.
For the parameters satisfying the above bound,
the function $f(r)$ has a single root that corresponds to the position of the unique event horizon.
For $\Lambda<\frac{m^2}{\beta}$,
the point $r=\sqrt{\frac{2\beta}{\beta\Lambda-m^2}}$ is not the curvature singularity.

For $m^2= -\beta\Lambda$ which for $\beta>0$ requires $\Lambda<0$,
$A_1(r)$ vanishes and 
the solution \eqref{nonflat} reduces to the Schwarzschild- anti- de Sitter  in GR
with the cosmological constant $\Lambda$, Eq. \eqref{sch_ds}.

 %%%%%%%%%%%%%%%%%%%%%%%%%%%%%%%%%%%%%%%%%
\section{The case of the vector field with the form of the Coulomb potential}
\label{sec:solutions2}

In this section, 
we consider the case that the temporal component of the vector field  $A_0(r)$
is given by the Coulomb potential as well as the constant term $P$, 
\begin{align}
\label{pq}
A_0(r)=P+\frac{Q}{r},
\end{align}
where the constant $Q$ corresponds to the electric charge.

For $m=\beta=0$ where the gauge symmetry is recovered,
the Reissner-Nortsr\"om -(anti-) de Sitter solution is obtained by  
\begin{subequations}
\label{reissner}
\begin{align}
f(r)&=h(r)=1-\frac{\Lambda}{3}r^2 -\frac{2M}{r}+\frac{Q^2}{2m_p^2 r^2},
\\
A_1(r)&=0.
\end{align}
\end{subequations}

\subsection{The stealth Schwarzschild solution}

First, we consider the case of $m=0$ and $\Lambda=0$.
As argued in Ref. \cite{Chagoya:2016aar},
only for $\beta=\frac{1}{4}$ the stealth Schwarzschild solution \eqref{eq:stealth}
is obtained by
\begin{subequations}
\label{stealth2}
\begin{align}
\label{stealth2_1}
f(r)&
=h(r)=1-\frac{2M}{r},
\\
\label{stealth2_2}
A_1(r)
&=
\pm
\frac{\sqrt{Q^2+2P\left(Q+MP\right)r}}
      {r}
\frac{1}{f(r)}.
\end{align}
\end{subequations}
The positivity inside the square root \eqref{stealth2_2} for an arbitrary $r$
requires $P\left(Q+MP\right)\geq 0$.

%%%%%%%%%%%%%%%%%%%%%%%%
\subsection{The Schwarzschild- (anti-) de Sitter  solution}

We then consider the case that $m^2\neq 0$ and $\Lambda\neq 0$,
where the generalization of the Schwarzschild- (anti-) de Sitter solution \eqref{eq:selftuned} is obtained only for $\beta=\frac{1}{4}$.

In the case of $m^2>0$,  
only for $P=\pm \frac{m_p}{\sqrt{2}m}  \sqrt{4m^2+\Lambda}$ in Eq. \eqref{pq}
the  Schwarzschild- anti- de Sitter  solution
is obtained by 
\begin{subequations}
\label{selftune2}
\begin{align}
f(r)&
=h(r)
=1-\frac{2M}{r}+\frac{4m^2}{3}r^2,
\\
\label{selftune2_c}
A_1(r)
&=\pm
\frac{1}{\sqrt{3} m r f(r)}
\left[
  m_p^2 r (3M- 2m^2 r^3) (4m^2+\Lambda)
\right.
\nonumber\\
&\left.
\pm 3 m m_p r\sqrt{2\left(4m^2+\Lambda\right)} Q
+3m^2 Q^2
\right]^{\frac{1}{2}}.
\end{align}
\end{subequations}
Similarly in the case of $m^2<0$,
only for $P=\pm \frac{m_p}{\sqrt{2}|m|}\sqrt{4|m|^2-\Lambda}$ in Eq. \eqref{pq},
the Schwarzschild- de Sitter solution is obtained by
\begin{subequations}
\label{selftune3}
\begin{align}
f(r)&
=h(r)
=1-\frac{2M}{r}-\frac{4|m|^2}{3}r^2,
\\
\label{selftune3_c}
A_1(r)
&=
\pm
\frac{1}{\sqrt{3}|m|rf(r)}
\left[
  m_p^2 r (3M+ 2|m|^2 r^3) (4|m|^2-\Lambda)
\right.
\nonumber \\
&\left.
\pm 3 |m| m_p r\sqrt{2\left(4|m|^2-\Lambda\right)} Q
+3|m|^2 Q^2
\right]^{\frac{1}{2}}.
\end{align}
\end{subequations}

For the solution \eqref{selftune2},
in order for $A_0(r)$ to be real, we require $\Lambda \geq -4m^2$.
For a large $r$, however, the combination inside the square root in Eq. \eqref{selftune2_c} 
always becomes negative, and hence the solution \eqref{selftune2} may not be regarded as the physical one.
On the other hand,  
for the solution \eqref{selftune3},
in order for $A_0(r)$ to be real, we require that $\Lambda <4|m|^2$
and then the positivity inside the square root of Eq. \eqref{selftune3_c}
between the event and cosmological horizons
can be naturally realized.

\subsection{The asymptotically anti- de Sitter  solution}

Finally, we consider the case of $P=0$ in Eq. \eqref{pq}.
As for the other cases, 
only for $\beta=\frac{1}{4}$,
the asymptotically anti- de Sitter solution \eqref{nonflat} is obtained by 
\begin{subequations}
\label{chagoya2}
\begin{align}
f(r)
&=
\frac{1}
{6mr (\Lambda-4m^2)^2}
\nonumber\\
&\times
 \left\{-6 \Lambda ^2 mr
+128 m^7 r^3-32 m^5
   \left(24 M+2 \Lambda  r^3-9 r\right)
\right.
\nonumber\\
&\left.
+8 \Lambda  m^3 r  \left(\Lambda  r^2-6\right)
+3 \left(\Lambda +4
   m^2\right)^2 {\rm arctan}(2 m r)
   \right\},
\\
h(r)&=\frac{\left(\Lambda -4 m^2\right)^2 \left(4 m^2 r^2+1\right)^2} 
              {16 m^4 \left(4 m^2 r^2-\Lambda r^2+2\right)^2}
f(r),
\\
\label{chagoya2_3}
A_1(r)
&=\pm 
\frac{\sqrt{Q^2\left(1+4m^2 r^2\right)-2m_p^2 \left(\Lambda +4m^2\right)r^4f(r)}}
    {r \sqrt{f(r)h(r)} \sqrt{1+4r^2m^2}}.
\end{align}
\end{subequations}
In order for $A_1(r)$ to be real for $f(r)>0$, 
$\Lambda+4m^2\leq 0$.
From the large-$r$ limit of the metric functions $f(r)$ and $h(r)$,
\begin{align}
f(r)&= \frac{4m^2}{3}r^2+\frac{12m^2+\Lambda}{4m^2-\Lambda} +{\cal O}\Big(\frac{1}{r}\Big),
\nonumber \\
h(r)&= \frac{4m^2}{3}r^2+\frac{28m^2+\Lambda}{12m^2-3\Lambda} +{\cal O}\Big(\frac{1}{r}\Big),
\end{align}
the effective cosmological constant is read as $\Lambda_{\rm eff}=-4m^2<0$
and hence the spacetime is asymptotically anti- de Sitter.
For $M>0$,
the function $f(r)$ has a single root which corresponds to the position of the unique event horizon.
For $\Lambda\leq 4m^2$,
the point $r=\sqrt{\frac{2}{\Lambda-4m^2}}$ is not the curvature singularity.

For $m=\pm \frac{\sqrt{-\Lambda}}{2}$
where the positivity of $m^2$ requires $\Lambda<0$,
the solution \eqref{chagoya2} reduces to the Schwarzschild- anti- de Sitter  
\begin{align}
\label{grlimit}
h(r)&=f(r)=1-\frac{2M}{r}-\frac{\Lambda}{3}r^2,
\nonumber\\
A_1(r)&=\pm \frac{Q}{rf}.
\end{align}
Equation \eqref{grlimit} also corresponds to the $m\to \pm \frac{\sqrt{-\Lambda}}{2}$ limit of 
Eqs. \eqref{selftune2} and \eqref{selftune3}.

\section{The other specific solutions}
\label{sec:others}

In this section, we consider the cases 
where the temporal component of the vector field 
is not given by Eq. \eqref{pq}.

\subsection{The solutions for the more general form of $A_0(r)$}

First, we consider the case 
where the additional (inverse) power-law function of the radial coordinate $r$ 
is added to $A_0(r)$ shown in Eq. \eqref{pq},
namely,
\begin{align}
\label{pqr}
A_0(r)&
=P+\frac{Q}{r}+Q_p r^p,
\end{align} 
where $p$ is the real number ($p\neq -1$)
and $Q_p$ is the constant.

For the temporal component of the vector field given by Eq. \eqref{pqr},
the existence of the solution for an arbitrary $p$
requires $\beta=\frac{1}{4}$, $P=\pm 2m_p$ and $m=\pm \frac{\sqrt{\Lambda}}{2}$ ($\Lambda>0$). 

For $p\neq -3, -\frac{3}{2}, -\frac{1}{2}$,
the solution is given by 
\begin{widetext}
\begin{subequations}
\label{chagoya3}
\begin{align}
f(r)&=
\frac{1}{4m_p^2 r}
\left[-8 M m_p^2+\frac{4}{3} \Lambda  m_p^2
   r^3+4 m_p^2 r
%+
\pm 4 m_p Q_p r^{p+1}
   \left(\frac{\Lambda  (p+1) r^2}{p+3}+1\right)+(p+1)^2
   Q_p^2 r^{2 p+1} \left(\frac{\Lambda  r^2}{2
   p+3}+\frac{1}{2 p+1}\right)\right],
\\
h(r)&=\frac{1}{ \left(1\pm \frac{(p+1) Q_pr^p}  {2m_p}\right)^2}f(r),
\\
A_1(r)
&=
\pm 
2  \sqrt{3} m_p
\sqrt{%4 p^3+20 p^2+27 p+9
(2p+1)(2p+3)(p+3)} 
 (\pm 2 m_p+(p+1)Q_p r^p)
\nonumber\\
&\times 
\left\{
-r 
\left[
r \left(
p^3 \big(16 \Lambda 
   m_p^2 r^2
\pm 
48 \Lambda  m_p Q_p
   r^{p+2}+
3 Q_p^2 r^{2 p} 
(11 \Lambda   r^2+9)
\big)
\right.
\right.
\right.
\nonumber\\
&\left.
 \left.
+p^2 
\big(80 \Lambda  m_p^2
   r^2
\pm 
144 \Lambda  m_p Q_p r^{p+2}
+3Q_p^2 r^{2 p} 
(19 \Lambda r^2+9)
\big)
+3 \Lambda  p r^2 
 (36m_p^2
\pm
 44 m_p Q_p r^p
+13 Q_p^2  r^{2 p})
\right.
\right.
\nonumber\\
&\left.
  \left.
+9 \Lambda  r^2
 (\pm 2m_p+Q_p r^p)^2
+6 p^4 Q_p^2  r^{2 p}
(\Lambda  r^2+1)
\right)
-24 M  m_p^2
 %(4 p^3+20 p^2+27 p+9)
(2p+1)(2p+3)(p+3)
\right]
\nonumber\\
&\left.
+6 %(4 p^3+20 p^2+27 p+9) 
(2p+1)(2p+3)(p+3)Q r 
(\pm 2 m_p+Q_p r^p)
+3 %(4 p^3+20 p^2+27  p+9) 
(2p+1)(2p+3)(p+3)Q^2
\right\}^{\frac{1}{2}}
\nonumber\\
&\times
\Big\{
r \left[4 m_p^2%(4 p^3+20 p^2+27 p+9) 
(2p+1)(2p+3)(p+3)(\Lambda r^2+3)
\pm 12 m_p% (4 p^2+8 p+3) 
(2p+1)(2p+3)Q_p r^p 
(\Lambda  p r^2+p+\Lambda  r^2+3)
\right.
\nonumber\\
&
\left.
+3 (p+1)^2 (p+3) Q_p^2 r^{2 p}
  (2 p (\Lambda  r^2+1)+\Lambda r^2+3)
\right]
-24 M m_p^2 
(2p+1)(2p+3)(p+3)
%(4 p^3+20  p^2+27 p+9)
\Big\}^{-1},
\end{align} 
\end{subequations}
\end{widetext}
where the upper and lower branches of Eq. \eqref{chagoya3}
correspond to $P=2m_p$ and $P=-2m_p$, respectively.
%%%%%
The point $r= \left[\mp \frac{2m_p}{(p+1)Q_p}\right]^{\frac{1}{p}}$
could be the curvature singularity other than at $r=0$.
For $P=2m_p$, 
the appearance of the curvature singularity can be avoided
for $Q_p<0$ and $p<-1$ or for $Q_p>0$ and $p>-1$,
while 
for $P=-2m_p$
it can be avoided 
for $Q_p>0$ and $p<-1$ or for $Q_p<0$ and $p>-1$.
%%%%%
For any value of $p(\neq -3,-\frac{3}{2},-\frac{1}{2})$ and $M>0$,
the function $f(r)$ has a single root
which corresponds to the position of the unique event horizon.
For a larger value of $M$,
the singularity $r= \left[\mp \frac{2m_p}{(p+1)Q_p}\right]^{\frac{1}{p}}$
is hidden by the event horizon.

For the other values of $p=-3,-\frac{3}{2},-\frac{1}{2}$,
the similar solutions are obtained only for $\beta=\frac{1}{4}$ and $P=\pm 2m_p$.
Here, we introduce the solution for each case:
\begin{widetext}
%%%%%%%%%%%%%%%%%%%%%%%%%%%%%%%
\begin{enumerate}

\item
For $p=-3$,
the solution for $P=\pm 2m_p$ is given by  
\begin{subequations}
\label{pm3}
\begin{align}
f(r)&=
\frac{1}{15m_p^2 r^6}
\left[
\pm 15 m_p Q_{-3} r^3
-Q_{-3}^2 (3+5r^2\Lambda)
+
 5m_p^2 r^5
  (-6M + 3 r + r^3\Lambda)
\mp 30 m_p Q_{-3} r^5\Lambda \ln (r)
\right],
\\
h(r)&=\frac{m_p^2 r^6}
               {\left(Q_{-3}\mp m_p r^3\right)^2}
      f(r),
\\
A_1(r)
&=
\pm
\frac{\sqrt{15} m_p (Q_{-3}\mp m_p r^3)}
{\pm 15 m_p Q_{-3} r^3
-Q_{-3}^2 (3+5r^2\Lambda)
+
 5m_p^2 r^5
  (-6M + 3 r + r^3\Lambda)
\mp 30 m_p Q_{-3} r^5\Lambda \ln (r)}
\nonumber\\
&\times
\Big\{
  30 Q Q_{-3} r^2
+ Q_{-3}^2
   \left(27+20r^2\Lambda\right)
+5 r^4
  \left(
   3Q^2 
\pm 12 \left(\pm 2Mm_p+Q\right) m_p r
-4m_p^2 r^4\Lambda
  \right)
\pm 120 m_p Q_{-3} r^5\Lambda \ln(r)
\Big\}^{\frac{1}{2}}.
\end{align}
\end{subequations}
The point $r=\left(\pm \frac{Q_{-3}}{m_p}\right)^{\frac{1}{3}}$ 
could be the curvature singularity other than at $r=0$,
which is absent for $Q_{-3}<0$ in the positive branch 
and  for $Q_{-3}>0$ in the negative branch.

\item
For $p=-\frac{3}{2}$,
the solution for $P=\pm 2m_p$ is given by
\begin{subequations}
\label{p32}
\begin{align}
f(r)
&=\frac{1}{96m_p^2 r^3}
\left[
-3 Q_{-3/2}^2
\mp 32m_p Q_{-3/2} r^{\frac{3}{2}}\left(-3+r^2\Lambda\right)
+32m_p^2 r^2\left(-6M + r(3+r^2\Lambda)\right)
+6 Q_{-3/2}^2 r^2\Lambda \ln (r)
\right],
\\
h(r)&=\frac{16m_p^2 r^3} 
              {\left(Q_{-3/2}\mp 4m_p r^{\frac{3}{2}}\right)^2}
f(r),
\\
A_1(r)
&=
\pm
\frac{2 \sqrt{6} m_p \left(Q_{-3/2}\mp 4 m_p r^{3/2}\right)}
       {-3 Q_{-3/2}^2
\mp 32m_p Q_{-3/2} r^{\frac{3}{2}}\left(-3+r^2\Lambda\right)
+32m_p^2r^2\left(-6M + r(3+r^2\Lambda)\right)
+6 Q_{-3/2}^2 r^2\Lambda \ln (r)}
\nonumber\\
&\times
\Big\{
8 r \big(%3 M r
24m_p^2M r
-4 \Lambda  m_p^2 r^4
 \pm 12m_p Q r
+3 Q^2\big)
+16Q_{-3/2} \sqrt{r} (\pm 2 \Lambda m_p r^3+3 Q)
-6 \Lambda  Q_{-3/2}^2 r^2 \ln (r)
+27 Q_{-3/2}^2
\Big\}^{\frac{1}{2}}.
\end{align}
\end{subequations}
The point $r=\left(\pm \frac{Q_{-3/2}}{4m_p}\right)^{\frac{2}{3}}$ 
could be the curvature singularity other than at $r=0$,
which is absent 
for $Q_{-3/2}<0$ in the positive branch 
and 
for $Q_{-3/2}>0$ in the negative branch.

\item
For $p=-\frac{1}{2}$ ,
the solution for $P=\pm 2m_p$ is given by 
\begin{subequations}
\label{p12}
\begin{align}
f(r)
&=\frac{1}{480 m_p^2 r}
\left[
-960 M m_p^2 
+15 Q_{-1/2}^2 r^2\Lambda
+160 m_p^2 r \left(3+r^2\Lambda\right)
\pm 96 m_p Q_{-1/2} \sqrt{r} \left(5+r^2\Lambda\right)
+30 Q_{-1/2}^2 \ln (r)
\right],
\\
h(r)&=\frac{16m_p^2 r} 
              {\left(Q_{-1/2}\pm 4m_p r^{\frac{1}{2}}\right)^2}
f(r),
\\
A_1(r)
&=\pm 
\frac{2 \sqrt{30} m_p \left(Q_{-1/2}\pm 4 m_p r^{\frac{1}{2}}\right)}
    { \sqrt{r}\left[
  -960 M m_p^2 
+15 Q_{-1/2}^2 r^2\Lambda
+160 m_p^2 r \left(3+r^2\Lambda\right)
\pm 96 m_p Q_{-1/2} \sqrt{r} \left(5+r^2\Lambda\right)
+30 Q_{-1/2}^2 \ln (r)
  \right]}
\nonumber\\
&\times
\Big\{
960 M m_p^2 r
-160 \Lambda  m_p^2 r^4
+240 Q (\pm 2 m_p r+Q_{-1/2}\sqrt{r})
\mp 96 \Lambda m_p Q_{-1/2} r^{7/2}
+120 Q^2
\nonumber\\
&-15 \Lambda  Q_{-1/2}^2 r^3
+120 Q_{-1/2}^2 r
-30 Q_{-1/2}^2 r \ln (r)
\Big\}^{\frac{1}{2}}.
\end{align}
\end{subequations}
The point $r=\left(\mp \frac{Q_{-1/2}}{4m_p}\right)^2$ 
could be the curvature singularity other than at $r=0$,
which is absent
for $Q_{-1/2}>0$ in the positive branch 
and 
for $Q_{-1/2}<0$ in the negative branch.

\end{enumerate}
\end{widetext}

%%%%%%%%%%%%%%%%%%%%%%%
For all the values of $p=-3,-\frac{3}{2}, -\frac{1}{2}$,
the functions $f(r)$ and $h(r)$ have a single root 
which corresponds to the position of the unique event horizon.
In the limit of $\Lambda\to 0$ and $Q_p\to 0$,
the solutions \eqref{chagoya3}, \eqref{pm3}, \eqref{p32} and \eqref{p12}
reduce to the stealth Schwarzschild solution \eqref{stealth2} with $P=\pm 2m_p$.
%%%%%

Moreover, 
as argued in Ref. \cite{Chagoya:2016aar},
only for $m=0$, $\beta=\frac{1}{4}$ and $p=2$ in Eq. \eqref{pqr},
the other type of the solution can be obtained for 
\begin{align}
Q_2=  \frac{2m_p^2 P\Lambda}{3(P^2-4m_p^2)}, 
\end{align}
given by 
\begin{subequations}
\label{chagoya}
\begin{align}
f(r)&=1-\frac{2M}{r}
+\frac{4m_p^2r^2\Lambda\left(5P^2+m_p^2\left(-20+3r^2\Lambda\right)\right)}
        {15\left(P^2-4m_p^2\right)^2},
\\
h(r)&=\frac{\left(P^2-4m_p^2\right)^2}
{\left(P^2+2m_p^2\left(-2+r^2\Lambda\right)\right)^2}
      f(r),
%\\
%A_0(r)&
%=P+\frac{Q}{r}
%+\frac{2m_p^2 P\Lambda r^2}
%        {3\left(P^2-4m_p^2\right)},
\\
A_1(r)
&=
\pm
\sqrt{\frac{A_0(r)^2}{f(r)h(r)}
-\frac{P^2+2m_p^2r^2\Lambda}{h(r)}}.
\end{align}
\end{subequations}
The spacetime is neither asymptotically Minkowski nor (anti-) de Sitter.
If $\frac{4m_p^2-P^2}{\Lambda}>0$,
the point $r=\frac{1}{m_p}\sqrt{\frac{4m_p^2-P^2}{2\Lambda}}$
is the curvature singularity
other than $r=0$.
On the other hand, 
if $\frac{4m_p^2-P^2}{\Lambda}<0$,
there is no curvature singularity except for $r=0$.
In both the cases, 
the function $f(r)$ always has a single root which corresponds to the position of the unique event horizon,
and for $\frac{4m_p^2-P^2}{\Lambda}>0$ the singularity at $r=\frac{1}{m_p}\sqrt{\frac{4m_p^2-P^2}{2\Lambda}}$
is hidden by the event horizon 
$\sqrt{\Lambda}M>\frac{4}{15m_p}\sqrt{\frac{4m_p^2-P^2}{2}}$.
For the other values of $p$
the analytic solution similar to Eq. \eqref{chagoya} could not be found.

\subsection{The solution for $A_1(r)=0$}

So far, we have investigated the static and spherically symmetric solutions for several choices of $A_0(r)$.
Instead, we may specify $A_1(r)$ and then find the other variables $f(r)$, $h(r)$ and $A_0(r)$
by solving the equations of motion \eqref{eom_a} and \eqref{eom_b}
under the ansatz \eqref{metric_ansatz} and \eqref{proca_ansatz}.
For $m\neq0$ and/or $\beta\neq 0$,
the $r$ component of the vector field equation of motion \eqref{eom_b} becomes nontrivial as 
\begin{align}
\label{Pr}
0&=
h(r) A_1(r)
\nonumber\\
&\times 
\left[
-m^2 r^2 f(r)
+\beta
\left(
-f(r)(1-h(r))+r h(r) f'(r)
\right)
\right],
\end{align}
which 
as in general $h(r)\neq 0$ 
gives the two possibilities
\begin{subequations}
\begin{align}
\label{eqs}
&-m^2 r^2f (r)
+\beta
\left(
-f(r)(1-h(r))+r h(r) f'(r)
\right)
=0,
\\
&
{\rm or}
\nonumber\\
\label{eqs2}
&A_1(r)=0.
\end{align}
\end{subequations}
The solutions, Eqs. \eqref{eq:stealth}, \eqref{eq:selftuned}, \eqref{nonflat}, \eqref{stealth2}, \eqref{selftune2}, \eqref{selftune3},
\eqref{chagoya2}, \eqref{chagoya3}, \eqref{pm3}, \eqref{p32}, \eqref{p12}, and \eqref{chagoya},
have been originated from the former choice \eqref{eqs}.
Under the ansatz Eqs. \eqref{metric_ansatz} and \eqref{proca_ansatz},
if the $r$ component of the vector field equation of motion \eqref{Pr} is satisfied 
then the $(t,r)$ component of the Einstein equation is also automatically satisfied.

For the latter case \eqref{eqs2},
the specific solution is obtained 
only for $m=0$, $\Lambda=0$ and $\beta=\frac{1}{4}$
by 
\begin{align}
\label{a10}
f(r)&=h(r)=1\pm \sqrt{\frac{r_0}{r}},
\quad
A_0(r)= 2 m_p f(r),
\end{align}
where $r_0$ is the integration constant.
The solution \eqref{a10} was obtained in Ref. \cite{Geng:2015kvs}.
The spacetime is asymptotically flat.
The singularity at $r=0$ is visible in the positive branch and hidden by the event horizon at $r=r_0$ in the negative branch,
respectively.
The solution similar to Eq. \eqref{a10}
could not be found for the more general cases
of $m\neq 0$, $\Lambda\neq 0$, or $\beta\neq \frac{1}{4}$.

\subsection{A short summary}
\label{sec:short_summary}

Throughout Secs. \ref{sec:solutions2} and \ref{sec:others},
we have obtained the static and spherically symmetric solutions for several nontrivial choices of 
the temporal component of the vector field $A_0(r)$.

For $A_0(r)$ with the form of the Coulomb potential given by Eq. \eqref{pq},
we have obtained the stealth Schwarzschild, the Schwarzschild- (anti-) de Sitter and the asymptotically anti- de Sitter  solutions 
\eqref{stealth2},  \eqref{selftune2} and \eqref{selftune3}, and \eqref{chagoya2},
respectively.
Unexpectedly,  
these solutions are present only
for the specific value of the nonmimal coupling constant, $\beta=\frac{1}{4}$,
and the electric charge $Q$ does not appear in the metric,
which is different from the case of the Reissner- Nortsr\"om [- (anti-) de Sitter] solution \eqref{reissner}.

For the other cases,
we could obtain the solutions
\eqref{chagoya3}, \eqref{pm3}, \eqref{p32}, \eqref{p12}, \eqref{chagoya} and \eqref{a10}.
All these solutions also exist only for $\beta=\frac{1}{4}$.

%%%%%%%%%%%%%%%%%%%%%%%%%%%%%%%
\section{The slowly rotating solutions}
\label{sec:slow}

Finally, we investigate the slowly rotating solutions
within the Hartle-Thorne approximation \cite{Hartle:1967he,Hartle:1968si},
where the rotational correction is obtained in the perturbation framework to the static and spherically symmetric background
with respect to the angular velocity of the black hole $\Omega$.

The correction to the static and spherically symmetric metric \eqref{metric_ansatz}
appears in the $(t,\phi)$ component at ${\cal O}(\Omega)$
and in the other components at  ${\cal O}(\Omega^2)$.
Thus within ${\cal O}(\Omega)$ the metric will take the form of 
\begin{align}
\label{eq:framedrag1}
ds^2&=-f(r)dt^2 +\frac{dr^2}{h(r)}
+r^2\left(d\theta^2+\sin^2\theta d\phi^2\right)
\nonumber\\
&-2r^2\omega(r) \sin^2\theta dtd\phi,
\end{align}
where $\omega(r)$ is the unknown function of ${\cal O}(\Omega)$.
Similarly, 
the correction to the vector field in the static and spherically symmetric background \eqref{proca_ansatz}
appears in the $\phi$ component at ${\cal O}(\Omega)$
and in the other components at ${\cal O} (\Omega^2)$.
Hence, 
within ${\cal O}(\Omega)$ the vector field will take the form of 
\begin{align}
A_\mu= \left(A_0(r), A_1(r),0,A_3(r,\theta)\right).
\end{align}
For the separability of the equations of motion at ${\cal O}(\Omega)$,
we assume that the azimuthal component of the vector field $A_3(r,\theta)$ takes the form 
\begin{align}
\label{eq:framedrag2}
A_3(r,\theta)&= a_3 (r)\sin^2\theta,
\end{align}
where $a_3(r)$ is the other unknown function of ${\cal O} (\Omega)$.

The unknown functions $\omega(r)$ and $a_3(r)$
in Eqs. \eqref{eq:framedrag1} and \eqref{eq:framedrag2}
are found as the solution of the field equations \eqref{eom_a} and \eqref{eom_b} at ${\cal O}(\Omega)$.
%%%%%%%%%%%%%%%%%%%%%%
At ${\cal O}(\Omega)$,
only the $(t,\phi)$ component of the Einstein equation \eqref{eom_a} becomes nontrivial,
and similarly
only the $\phi$ component of the vector field equation of motion \eqref{eom_b} 
becomes nontrivial.
%%%%%%%%%%%%%%%%%%%%%
Thus they will be solved
under the boundary conditions that $\omega(r)$ and $a_3(r)$ are finite in the large-$r$ limit.

Before starting,
we consider the case of $m=\beta=0$ where the gauge symmetry is recovered.
In this case,
the slow-rotation correction to the Reissner-Nortsr\"om-(anti-) de Sitter solution \eqref{reissner}
is obtained by
\begin{subequations}
\label{kerr_newman}
\begin{align}
\label{kn1}
\omega(r)
&=\omega_0+\frac{2J}{r^3}-\frac{JQ^2}{2m_p^2 M r^4},
\\
\label{kn2}
a_3(r)&=-\frac{JQ}{Mr},
\end{align}
\end{subequations}
which agrees with the Kerr-Newman-(anti-) de Sitter solution 
by neglecting the terms of ${\cal O} (\Omega^2)$ \cite{Dehghani:2002nt,Huaifan:2009nf}.

\subsection{For the background with the constant $A_0(r)$}

First, we consider the background solutions with the constant temporal component 
of the vector field, 
$A_0(r)=P$ discussed in Sec. \ref{sec:solutions},
namely,
Eqs. \eqref{eq:stealth} and \eqref{eq:selftuned} for $P\neq 0$
and Eq. \eqref{nonflat} for $P= 0$.

For these background solutions,
if we assume that $a_3(r)=0$,
we find that
$\omega(r)$ remains the same as the slow-rotation limit of the Schwarzschild- (anti-) de Sitter background in GR,
given by 
\begin{align}
\label{sol:framedrag}
\omega (r)= \omega_0+\frac{2J}{r^3},
\end{align}
where the constant $\omega_0=0$ for the Schwarzschild background 
and $\omega_0\neq 0$ for the Schwarzschild- (anti-) de Sitter background,
and the constant $J$ represents the angular momentum of the black hole.
This is the confirmation of the argument in Sec. \ref{sec:solutions} at the level of the first order in the slow-rotation approximation, ${\cal O} (\Omega)$,
as the solutions in the scalar-tensor theory \eqref{eq:action3} obtained in Refs. \cite{Cisterna:2015uya,Maselli:2015yva}
are expressed as those in the generalized Proca theory with the vanishing field strength.

On the other hand,
if we assume that $a_3(r)\neq 0$,
formally
the more general solution can be found.
For instance,
for the stealth Schwarzschild background \eqref{eq:stealth} with $P =\pm \frac{m_p}{\sqrt{\beta}}$ ($\beta>0$),
the solution is given by 
\begin{subequations} 
\label{nonzeroa2}
\begin{align}
\label{omega2}
\omega(r)
&=
\omega_0
+\frac{2J}{r^3}
\mp \frac{3{\cal Q} M}{2m_p \sqrt{\beta} r^4},
\\ 
\label{a32}
a_3(r)&=\frac{\cal Q}{r},
\end{align} 
\end{subequations}
where ${\cal Q}$ is the integration constant.
The same solution as Eq. \eqref{nonzeroa2}
is also obtained for the Schwarzschild- (anti-) de Sitter  background \eqref{eq:selftuned}
$m=\pm\sqrt{\beta\Lambda}$
($\beta \Lambda>0$).
For the other background parameters,
we could obtain the solutions with the same leading behavior 
as Eq. \eqref{nonzeroa2}
in the large-$r$ limit.
From Eq. \eqref{omega2}, 
the contribution of ${\cal Q}$ seems to appear 
as an independent ``charge''.

In fact,
if we consider 
the slow-rotation correction to the 
Schwarzschild- (anti-) de Sitter solution with $m=\beta=0$
by assuming that $a_3(r)\neq 0$,
the solution
with the same leading behavior in the large-$r$ limit as Eq. \eqref{a32}
could be obtained.
However, 
we have to set the integration constant ${\cal Q}=0$,
as the slow-rotation correction
to the electrically neutral background solution
could not induce the nonzero magnetic field.
Similarly in our case,
for the background 
with the vanishing electric field strength
it is reasonable to set ${\cal Q}=0$,
and hence 
Eq. \eqref{sol:framedrag} with $a_3(r)=0$
can be regarded as the only physical solution.

\subsection{For the background with the nonconstant $A_0(r)$}

Second, we consider the background solutions with the nonconstant $A_0(r)$
discussed in Sec. \ref{sec:solutions2},
namely Eqs. \eqref{stealth2}, \eqref{selftune2}, \eqref{selftune3} 
and \eqref{chagoya2} with $m=\pm \frac{\sqrt{-\Lambda}}{2}$ where the solution reduces to Eq. \eqref{grlimit}.
We find that  $\omega(r)$ is given by Eq. \eqref{sol:framedrag} which is the same as the Kerr- (anti-) de Sitter solution,
and $a_3(r)$ is given by 
\begin{align}
\label{a33}
a_3(r)=-\frac{JQ}{M r},
\end{align}
which is similar to the result obtained in Ref. \cite{Chagoya:2016aar} for the stealth Schwarzschild background \eqref{stealth2}.
Thus $a_3(r)$ is the same 
as the slow-rotation limit of the Kerr-Newman-(anti-) de Sitter solution \eqref{kn2},
but $\omega(r)$ does not contain the term depending on the background electric charge $Q$.
This is the stealth property realized at the first order in the slow-rotation approximation, ${\cal O} (\Omega)$.
For Eq. \eqref{chagoya2} with $m\neq \pm \frac{\sqrt{-\Lambda}}{2}$,
no analytic solution of $\omega(r)$ and $a_3(r)$ could be obtained.

\subsection{A short summary}

In this section, we have investigated the slow-rotation corrections to the static and
spherically symmetric backgrounds
within the first order of the Hartle-Thorne approximation \cite{Hartle:1967he,Hartle:1968si}.

For the background with the vanishing electric field strength,
the slow-rotation correction to the metric was found to be the same as
the Kerr- (anti-) de Sitter solution \eqref{sol:framedrag}
with $A_3(r,\theta)=0$.

On the other hand,
for the background with the nonvanishing electric field strength,
the slow-rotation correction to the metric remains the same as the Kerr- (anti-) de Sitter solution \eqref{sol:framedrag},
but the azimuthal component $A_3(r,\theta)$ is the same as the Kerr- Newman- (anti-) de Sitter solution \eqref{a33},
which is the realization of the stealth property in the slowly rotating case.

%%%%%%%%%%%%%%%%%%%%%%%%%%%%%%
\section{Conclusions}
\label{sec:conclusions}

We have investigated the static and spherically symmetric solutions in the generalized Proca theory 
with the nonminimal coupling of the vector field to the Einstein tensor \eqref{eq:action}.
First, 
we have shown that the solutions obtained in the scalar-tensor theory with 
the nonminimal derivative coupling to the Einstein tensor \eqref{eq:action3} 
are also those in the generalized Proca theory \eqref{eq:action}
with the vanishing field strength,
and we have obtained the expressions of
the stealth Schwarzschild, 
the Schwarzschild- (anti-) de Sitter 
and 
the asymptotically anti- de Sitter  solutions in the generalized Proca theory.

Second, we have investigated these solutions  
where the temporal component of the vector field contains the term of the Coulomb potential.
In this case, 
as argued in Ref. \cite{Chagoya:2016aar},
the extension of these solutions requires the special value of the nonminimal coupling parameter,
irrespective of the value of the mass term of the vector field and the asymptotic property of the spacetime.
We have also obtained the other nontrivial solutions for the same value of the coupling constant.

Finally, 
we have investigated the first-order slow-rotation corrections to the static and spherically symmetric solutions.
We have found that
for the background with the vanishing electric field strength
the slowly rotating solution remains the same as in GR.
For the background with the nonvanishing electric field strength,
the slow-rotation correction to the metric 
does not depend on the electric charge
and may be regarded as the realization of the stealth property in the context of the slow-rotation approximation.

There will be various extensions of the present work.
The first subject is to investigate the stability of the solutions obtained in this paper.
The stability of the black hole solutions in the scalar-tensor Horndeski theory \eqref{eq:action3}
has been investigated in Refs. \cite{Kobayashi:2012kh,Kobayashi:2014wsa,Cisterna:2015uya,Ogawa:2015pea}.
Especially in Ref. \cite{Ogawa:2015pea} it was argued that
the static and spherically symmetric black hole solutions with 
the constant canonical kinetic term $X_\varphi=-\frac{1}{2}g^{\mu\nu}\partial_\mu\varphi\partial_\nu\varphi$
are generically unstable in the vicinity of the event horizon.
It will be interesting
to investigate whether there is the same kind of instability in the vector-tensor theory.
The other interesting issue is 
to investigate the rapidly rotating black hole solutions
and the spectrum of the quasinormal modes,
which could make the deviations from the GR solutions more evident.
Other than the vacuum solutions, 
it will also be very important
to investigate the solutions of the neutron stars and the other compact objects.
(See, e.g., \cite{Cisterna:2015yla,Cisterna:2016vdx,Maselli:2016gxk,Brihaye:2016lin} for the related studies in the case of the scalar-tensor Horndeski theory.)
We hope to come back to these issues in future work.

%%%%%%%%%%%%%%%%%%%%%%%%
\section*{Acknowledgements}
We thank E. Babichev, M. Kimura and H. O. Silva for comments.
This work was supported by FCT-Portugal through Grant No. SFRH/BPD/88299/2012. 
%%%%%%%%%%%%%%%%%%%%%%%

%%%%%%%%%%%%%%%%%%%%%%%%%%%%%%%%%%%%%%%%%%%
\bibliography{bibmonster} 

%merlin.mbs apsrev4-1.bst 2010-07-25 4.21a (PWD, AO, DPC) hacked
%Control: key (0)
%Control: author (0) dotless jnrlst
%Control: editor formatted (1) identically to author
%Control: production of article title (0) allowed
%Control: page (1) range
%Control: year (0) verbatim
%Control: production of eprint (0) enabled
\begin{thebibliography}{53}%
\makeatletter
\providecommand \@ifxundefined [1]{%
 \@ifx{#1\undefined}
}%
\providecommand \@ifnum [1]{%
 \ifnum #1\expandafter \@firstoftwo
 \else \expandafter \@secondoftwo
 \fi
}%
\providecommand \@ifx [1]{%
 \ifx #1\expandafter \@firstoftwo
 \else \expandafter \@secondoftwo
 \fi
}%
\providecommand \natexlab [1]{#1}%
\providecommand \enquote  [1]{``#1''}%
\providecommand \bibnamefont  [1]{#1}%
\providecommand \bibfnamefont [1]{#1}%
\providecommand \citenamefont [1]{#1}%
\providecommand \href@noop [0]{\@secondoftwo}%
\providecommand \href [0]{\begingroup \@sanitize@url \@href}%
\providecommand \@href[1]{\@@startlink{#1}\@@href}%
\providecommand \@@href[1]{\endgroup#1\@@endlink}%
\providecommand \@sanitize@url [0]{\catcode `\\12\catcode `\$12\catcode
  `\&12\catcode `\#12\catcode `\^12\catcode `\_12\catcode `\%12\relax}%
\providecommand \@@startlink[1]{}%
\providecommand \@@endlink[0]{}%
\providecommand \url  [0]{\begingroup\@sanitize@url \@url }%
\providecommand \@url [1]{\endgroup\@href {#1}{\urlprefix }}%
\providecommand \urlprefix  [0]{URL }%
\providecommand \Eprint [0]{\href }%
\providecommand \doibase [0]{http://dx.doi.org/}%
\providecommand \selectlanguage [0]{\@gobble}%
\providecommand \bibinfo  [0]{\@secondoftwo}%
\providecommand \bibfield  [0]{\@secondoftwo}%
\providecommand \translation [1]{[#1]}%
\providecommand \BibitemOpen [0]{}%
\providecommand \bibitemStop [0]{}%
\providecommand \bibitemNoStop [0]{.\EOS\space}%
\providecommand \EOS [0]{\spacefactor3000\relax}%
\providecommand \BibitemShut  [1]{\csname bibitem#1\endcsname}%
\let\auto@bib@innerbib\@empty
%</preamble>
\bibitem [{\citenamefont {Clifton}\ \emph {et~al.}(2012)\citenamefont
  {Clifton}, \citenamefont {Ferreira}, \citenamefont {Padilla},\ and\
  \citenamefont {Skordis}}]{Clifton:2011jh}%
  \BibitemOpen
  \bibfield  {author} {\bibinfo {author} {\bibfnamefont {Timothy}\ \bibnamefont
  {Clifton}}, \bibinfo {author} {\bibfnamefont {Pedro~G.}\ \bibnamefont
  {Ferreira}}, \bibinfo {author} {\bibfnamefont {Antonio}\ \bibnamefont
  {Padilla}}, \ and\ \bibinfo {author} {\bibfnamefont {Constantinos}\
  \bibnamefont {Skordis}},\ }\bibfield  {title} {\enquote {\bibinfo {title}
  {{Modified Gravity and Cosmology}},}\ }\href {\doibase
  10.1016/j.physrep.2012.01.001} {\bibfield  {journal} {\bibinfo  {journal}
  {Phys.Rept.}\ }\textbf {\bibinfo {volume} {513}},\ \bibinfo {pages} {1--189}
  (\bibinfo {year} {2012})},\ \Eprint {http://arxiv.org/abs/1106.2476}
  {arXiv:1106.2476 [astro-ph.CO]} \BibitemShut {NoStop}%
%%CITATION = ARXIV:1106.2476;%%
\bibitem [{\citenamefont {Berti}\ \emph {et~al.}(2015)\citenamefont {Berti}
  \emph {et~al.}}]{Berti:2015itd}%
  \BibitemOpen
  \bibfield  {author} {\bibinfo {author} {\bibfnamefont {Emanuele}\
  \bibnamefont {Berti}} \emph {et~al.},\ }\bibfield  {title} {\enquote
  {\bibinfo {title} {{Testing General Relativity with Present and Future
  Astrophysical Observations}},}\ }\href {\doibase
  10.1088/0264-9381/32/24/243001} {\bibfield  {journal} {\bibinfo  {journal}
  {Class. Quant. Grav.}\ }\textbf {\bibinfo {volume} {32}},\ \bibinfo {pages}
  {243001} (\bibinfo {year} {2015})},\ \Eprint
  {http://arxiv.org/abs/1501.07274} {arXiv:1501.07274 [gr-qc]} \BibitemShut
  {NoStop}%
%%CITATION = ARXIV:1501.07274;%%
\bibitem [{\citenamefont {Fujii}\ and\ \citenamefont
  {Maeda}(2007)}]{Fujii:2003pa}%
  \BibitemOpen
  \bibfield  {author} {\bibinfo {author} {\bibfnamefont {Y.}~\bibnamefont
  {Fujii}}\ and\ \bibinfo {author} {\bibfnamefont {K.}~\bibnamefont {Maeda}},\
  }\href {http://www.cambridge.org/uk/catalogue/catalogue.asp?isbn=0521811597}
  {\emph {\bibinfo {title} {{The scalar-tensor theory of gravitation}}}}\
  (\bibinfo  {publisher} {Cambridge University Press},\ \bibinfo {year}
  {2007})\BibitemShut {NoStop}%
%%CITATION = INSPIRE-618647;%%
\bibitem [{\citenamefont {Woodard}(2015)}]{Woodard:2015zca}%
  \BibitemOpen
  \bibfield  {author} {\bibinfo {author} {\bibfnamefont {Richard~P.}\
  \bibnamefont {Woodard}},\ }\bibfield  {title} {\enquote {\bibinfo {title}
  {{Ostrogradsky's theorem on Hamiltonian instability}},}\ }\href {\doibase
  10.4249/scholarpedia.32243} {\bibfield  {journal} {\bibinfo  {journal}
  {Scholarpedia}\ }\textbf {\bibinfo {volume} {10}},\ \bibinfo {pages} {32243}
  (\bibinfo {year} {2015})},\ \Eprint {http://arxiv.org/abs/1506.02210}
  {arXiv:1506.02210 [hep-th]} \BibitemShut {NoStop}%
%%CITATION = ARXIV:1506.02210;%%
\bibitem [{\citenamefont {Vainshtein}(1972)}]{Vainshtein:1972sx}%
  \BibitemOpen
  \bibfield  {author} {\bibinfo {author} {\bibfnamefont {A.I.}\ \bibnamefont
  {Vainshtein}},\ }\bibfield  {title} {\enquote {\bibinfo {title} {{To the
  problem of nonvanishing gravitation mass}},}\ }\href {\doibase
  10.1016/0370-2693(72)90147-5} {\bibfield  {journal} {\bibinfo  {journal}
  {Phys.Lett.}\ }\textbf {\bibinfo {volume} {B39}},\ \bibinfo {pages}
  {393--394} (\bibinfo {year} {1972})}\BibitemShut {NoStop}%
%%CITATION = PHLTA,B39,393;%%
\bibitem [{\citenamefont {Brax}\ \emph {et~al.}(2004)\citenamefont {Brax},
  \citenamefont {van~de Bruck}, \citenamefont {Davis}, \citenamefont {Khoury},\
  and\ \citenamefont {Weltman}}]{Brax:2004qh}%
  \BibitemOpen
  \bibfield  {author} {\bibinfo {author} {\bibfnamefont {Philippe}\
  \bibnamefont {Brax}}, \bibinfo {author} {\bibfnamefont {Carsten}\
  \bibnamefont {van~de Bruck}}, \bibinfo {author} {\bibfnamefont
  {Anne-Christine}\ \bibnamefont {Davis}}, \bibinfo {author} {\bibfnamefont
  {Justin}\ \bibnamefont {Khoury}}, \ and\ \bibinfo {author} {\bibfnamefont
  {Amanda}\ \bibnamefont {Weltman}},\ }\bibfield  {title} {\enquote {\bibinfo
  {title} {{Detecting dark energy in orbit - The Cosmological chameleon}},}\
  }\href {\doibase 10.1103/PhysRevD.70.123518} {\bibfield  {journal} {\bibinfo
  {journal} {Phys. Rev.}\ }\textbf {\bibinfo {volume} {D70}},\ \bibinfo {pages}
  {123518} (\bibinfo {year} {2004})},\ \Eprint
  {http://arxiv.org/abs/astro-ph/0408415} {arXiv:astro-ph/0408415 [astro-ph]}
  \BibitemShut {NoStop}%
%%CITATION = ASTRO-PH/0408415;%%
\bibitem [{\citenamefont {Horndeski}(1974)}]{Horndeski:1974wa}%
  \BibitemOpen
  \bibfield  {author} {\bibinfo {author} {\bibfnamefont {Gregory~Walter}\
  \bibnamefont {Horndeski}},\ }\bibfield  {title} {\enquote {\bibinfo {title}
  {{Second-order scalar-tensor field equations in a four-dimensional space}},}\
  }\href {\doibase 10.1007/BF01807638} {\bibfield  {journal} {\bibinfo
  {journal} {Int.J.Theor.Phys.}\ }\textbf {\bibinfo {volume} {10}},\ \bibinfo
  {pages} {363--384} (\bibinfo {year} {1974})}\BibitemShut {NoStop}%
%%CITATION = IJTPB,10,363;%%
\bibitem [{\citenamefont {Deffayet}\ \emph {et~al.}(2009)\citenamefont
  {Deffayet}, \citenamefont {Deser},\ and\ \citenamefont
  {Esposito-Far\`ese}}]{Deffayet:2009mn}%
  \BibitemOpen
  \bibfield  {author} {\bibinfo {author} {\bibfnamefont {C.}~\bibnamefont
  {Deffayet}}, \bibinfo {author} {\bibfnamefont {S.}~\bibnamefont {Deser}}, \
  and\ \bibinfo {author} {\bibfnamefont {G.}~\bibnamefont
  {Esposito-Far\`ese}},\ }\bibfield  {title} {\enquote {\bibinfo {title}
  {{Generalized Galileons: All scalar models whose curved background extensions
  maintain second-order field equations and stress-tensors}},}\ }\href
  {\doibase 10.1103/PhysRevD.80.064015} {\bibfield  {journal} {\bibinfo
  {journal} {Phys. Rev.}\ }\textbf {\bibinfo {volume} {D80}},\ \bibinfo {pages}
  {064015} (\bibinfo {year} {2009})},\ \Eprint {http://arxiv.org/abs/0906.1967}
  {arXiv:0906.1967 [gr-qc]} \BibitemShut {NoStop}%
%%CITATION = ARXIV:0906.1967;%%
\bibitem [{\citenamefont {Deffayet}\ \emph {et~al.}(2011)\citenamefont
  {Deffayet}, \citenamefont {Gao}, \citenamefont {Steer},\ and\ \citenamefont
  {Zahariade}}]{Deffayet:2011gz}%
  \BibitemOpen
  \bibfield  {author} {\bibinfo {author} {\bibfnamefont {C.}~\bibnamefont
  {Deffayet}}, \bibinfo {author} {\bibfnamefont {Xian}\ \bibnamefont {Gao}},
  \bibinfo {author} {\bibfnamefont {D.A.}\ \bibnamefont {Steer}}, \ and\
  \bibinfo {author} {\bibfnamefont {G.}~\bibnamefont {Zahariade}},\ }\bibfield
  {title} {\enquote {\bibinfo {title} {{From k-essence to generalised
  Galileons}},}\ }\href {\doibase 10.1103/PhysRevD.84.064039} {\bibfield
  {journal} {\bibinfo  {journal} {Phys.Rev.}\ }\textbf {\bibinfo {volume}
  {D84}},\ \bibinfo {pages} {064039} (\bibinfo {year} {2011})},\ \Eprint
  {http://arxiv.org/abs/1103.3260} {arXiv:1103.3260 [hep-th]} \BibitemShut
  {NoStop}%
%%CITATION = ARXIV:1103.3260;%%
\bibitem [{\citenamefont {Kobayashi}\ \emph {et~al.}(2011)\citenamefont
  {Kobayashi}, \citenamefont {Yamaguchi},\ and\ \citenamefont
  {Yokoyama}}]{Kobayashi:2011nu}%
  \BibitemOpen
  \bibfield  {author} {\bibinfo {author} {\bibfnamefont {Tsutomu}\ \bibnamefont
  {Kobayashi}}, \bibinfo {author} {\bibfnamefont {Masahide}\ \bibnamefont
  {Yamaguchi}}, \ and\ \bibinfo {author} {\bibfnamefont {Jun'ichi}\
  \bibnamefont {Yokoyama}},\ }\bibfield  {title} {\enquote {\bibinfo {title}
  {{Generalized G-inflation: Inflation with the most general second-order field
  equations}},}\ }\href {\doibase 10.1143/PTP.126.511} {\bibfield  {journal}
  {\bibinfo  {journal} {Prog. Theor. Phys.}\ }\textbf {\bibinfo {volume}
  {126}},\ \bibinfo {pages} {511--529} (\bibinfo {year} {2011})},\ \Eprint
  {http://arxiv.org/abs/1105.5723} {arXiv:1105.5723 [hep-th]} \BibitemShut
  {NoStop}%
%%CITATION = ARXIV:1105.5723;%%
\bibitem [{\citenamefont {Deffayet}\ \emph {et~al.}(2014)\citenamefont
  {Deffayet}, \citenamefont {G{\"u}mr{\"u}k{\c c}{\"u}o{\u g}lu}, \citenamefont
  {Mukohyama},\ and\ \citenamefont {Wang}}]{Deffayet:2013tca}%
  \BibitemOpen
  \bibfield  {author} {\bibinfo {author} {\bibfnamefont {Cedric}\ \bibnamefont
  {Deffayet}}, \bibinfo {author} {\bibfnamefont {A.~Emir}\ \bibnamefont
  {G{\"u}mr{\"u}k{\c c}{\"u}o{\u g}lu}}, \bibinfo {author} {\bibfnamefont
  {Shinji}\ \bibnamefont {Mukohyama}}, \ and\ \bibinfo {author} {\bibfnamefont
  {Yi}~\bibnamefont {Wang}},\ }\bibfield  {title} {\enquote {\bibinfo {title}
  {{A no-go theorem for generalized vector Galileons on flat spacetime}},}\
  }\href {\doibase 10.1007/JHEP04(2014)082} {\bibfield  {journal} {\bibinfo
  {journal} {JHEP}\ }\textbf {\bibinfo {volume} {04}},\ \bibinfo {pages} {082}
  (\bibinfo {year} {2014})},\ \Eprint {http://arxiv.org/abs/1312.6690}
  {arXiv:1312.6690 [hep-th]} \BibitemShut {NoStop}%
%%CITATION = ARXIV:1312.6690;%%
\bibitem [{\citenamefont {Tasinato}(2014)}]{Tasinato:2014eka}%
  \BibitemOpen
  \bibfield  {author} {\bibinfo {author} {\bibfnamefont {Gianmassimo}\
  \bibnamefont {Tasinato}},\ }\bibfield  {title} {\enquote {\bibinfo {title}
  {{Cosmic Acceleration from Abelian Symmetry Breaking}},}\ }\href {\doibase
  10.1007/JHEP04(2014)067} {\bibfield  {journal} {\bibinfo  {journal} {JHEP}\
  }\textbf {\bibinfo {volume} {04}},\ \bibinfo {pages} {067} (\bibinfo {year}
  {2014})},\ \Eprint {http://arxiv.org/abs/1402.6450} {arXiv:1402.6450
  [hep-th]} \BibitemShut {NoStop}%
%%CITATION = ARXIV:1402.6450;%%
\bibitem [{\citenamefont {Heisenberg}(2014)}]{Heisenberg:2014rta}%
  \BibitemOpen
  \bibfield  {author} {\bibinfo {author} {\bibfnamefont {Lavinia}\ \bibnamefont
  {Heisenberg}},\ }\bibfield  {title} {\enquote {\bibinfo {title}
  {{Generalization of the Proca Action}},}\ }\href {\doibase
  10.1088/1475-7516/2014/05/015} {\bibfield  {journal} {\bibinfo  {journal}
  {JCAP}\ }\textbf {\bibinfo {volume} {1405}},\ \bibinfo {pages} {015}
  (\bibinfo {year} {2014})},\ \Eprint {http://arxiv.org/abs/1402.7026}
  {arXiv:1402.7026 [hep-th]} \BibitemShut {NoStop}%
%%CITATION = ARXIV:1402.7026;%%
\bibitem [{\citenamefont {Allys}\ \emph {et~al.}(2016)\citenamefont {Allys},
  \citenamefont {Peter},\ and\ \citenamefont {Rodriguez}}]{Allys:2015sht}%
  \BibitemOpen
  \bibfield  {author} {\bibinfo {author} {\bibfnamefont {Erwan}\ \bibnamefont
  {Allys}}, \bibinfo {author} {\bibfnamefont {Patrick}\ \bibnamefont {Peter}},
  \ and\ \bibinfo {author} {\bibfnamefont {Yeinzon}\ \bibnamefont
  {Rodriguez}},\ }\bibfield  {title} {\enquote {\bibinfo {title} {{Generalized
  Proca action for an Abelian vector field}},}\ }\href {\doibase
  10.1088/1475-7516/2016/02/004} {\bibfield  {journal} {\bibinfo  {journal}
  {JCAP}\ }\textbf {\bibinfo {volume} {1602}},\ \bibinfo {pages} {004}
  (\bibinfo {year} {2016})},\ \Eprint {http://arxiv.org/abs/1511.03101}
  {arXiv:1511.03101 [hep-th]} \BibitemShut {NoStop}%
%%CITATION = ARXIV:1511.03101;%%
\bibitem [{\citenamefont {Beltran~Jimenez}\ and\ \citenamefont
  {Heisenberg}(2016)}]{Jimenez:2016isa}%
  \BibitemOpen
  \bibfield  {author} {\bibinfo {author} {\bibfnamefont {Jose}\ \bibnamefont
  {Beltran~Jimenez}}\ and\ \bibinfo {author} {\bibfnamefont {Lavinia}\
  \bibnamefont {Heisenberg}},\ }\bibfield  {title} {\enquote {\bibinfo {title}
  {{Derivative self-interactions for a massive vector field}},}\ }\href
  {\doibase 10.1016/j.physletb.2016.04.017} {\bibfield  {journal} {\bibinfo
  {journal} {Phys. Lett.}\ }\textbf {\bibinfo {volume} {B757}},\ \bibinfo
  {pages} {405--411} (\bibinfo {year} {2016})},\ \Eprint
  {http://arxiv.org/abs/1602.03410} {arXiv:1602.03410 [hep-th]} \BibitemShut
  {NoStop}%
%%CITATION = ARXIV:1602.03410;%%
\bibitem [{\citenamefont {De~Felice}\ \emph
  {et~al.}(2016{\natexlab{a}})\citenamefont {De~Felice}, \citenamefont
  {Heisenberg}, \citenamefont {Kase}, \citenamefont {Tsujikawa}, \citenamefont
  {Zhang},\ and\ \citenamefont {Zhao}}]{DeFelice:2016cri}%
  \BibitemOpen
  \bibfield  {author} {\bibinfo {author} {\bibfnamefont {Antonio}\ \bibnamefont
  {De~Felice}}, \bibinfo {author} {\bibfnamefont {Lavinia}\ \bibnamefont
  {Heisenberg}}, \bibinfo {author} {\bibfnamefont {Ryotaro}\ \bibnamefont
  {Kase}}, \bibinfo {author} {\bibfnamefont {Shinji}\ \bibnamefont
  {Tsujikawa}}, \bibinfo {author} {\bibfnamefont {Ying-li}\ \bibnamefont
  {Zhang}}, \ and\ \bibinfo {author} {\bibfnamefont {Gong-Bo}\ \bibnamefont
  {Zhao}},\ }\bibfield  {title} {\enquote {\bibinfo {title} {{Screening fifth
  forces in generalized Proca theories}},}\ }\href {\doibase
  10.1103/PhysRevD.93.104016} {\bibfield  {journal} {\bibinfo  {journal} {Phys.
  Rev.}\ }\textbf {\bibinfo {volume} {D93}},\ \bibinfo {pages} {104016}
  (\bibinfo {year} {2016}{\natexlab{a}})},\ \Eprint
  {http://arxiv.org/abs/1602.00371} {arXiv:1602.00371 [gr-qc]} \BibitemShut
  {NoStop}%
%%CITATION = ARXIV:1602.00371;%%
\bibitem [{\citenamefont {De~Felice}\ \emph
  {et~al.}(2016{\natexlab{b}})\citenamefont {De~Felice}, \citenamefont
  {Heisenberg}, \citenamefont {Kase}, \citenamefont {Mukohyama}, \citenamefont
  {Tsujikawa},\ and\ \citenamefont {Zhang}}]{DeFelice:2016yws}%
  \BibitemOpen
  \bibfield  {author} {\bibinfo {author} {\bibfnamefont {Antonio}\ \bibnamefont
  {De~Felice}}, \bibinfo {author} {\bibfnamefont {Lavinia}\ \bibnamefont
  {Heisenberg}}, \bibinfo {author} {\bibfnamefont {Ryotaro}\ \bibnamefont
  {Kase}}, \bibinfo {author} {\bibfnamefont {Shinji}\ \bibnamefont
  {Mukohyama}}, \bibinfo {author} {\bibfnamefont {Shinji}\ \bibnamefont
  {Tsujikawa}}, \ and\ \bibinfo {author} {\bibfnamefont {Ying-li}\ \bibnamefont
  {Zhang}},\ }\bibfield  {title} {\enquote {\bibinfo {title} {{Cosmology in
  generalized Proca theories}},}\ }\href {\doibase
  10.1088/1475-7516/2016/06/048} {\bibfield  {journal} {\bibinfo  {journal}
  {JCAP}\ }\textbf {\bibinfo {volume} {1606}},\ \bibinfo {pages} {048}
  (\bibinfo {year} {2016}{\natexlab{b}})},\ \Eprint
  {http://arxiv.org/abs/1603.05806} {arXiv:1603.05806 [gr-qc]} \BibitemShut
  {NoStop}%
%%CITATION = ARXIV:1603.05806;%%
\bibitem [{\citenamefont {De~Felice}\ \emph
  {et~al.}(2016{\natexlab{c}})\citenamefont {De~Felice}, \citenamefont
  {Heisenberg}, \citenamefont {Kase}, \citenamefont {Mukohyama}, \citenamefont
  {Tsujikawa},\ and\ \citenamefont {Zhang}}]{DeFelice:2016uil}%
  \BibitemOpen
  \bibfield  {author} {\bibinfo {author} {\bibfnamefont {Antonio}\ \bibnamefont
  {De~Felice}}, \bibinfo {author} {\bibfnamefont {Lavinia}\ \bibnamefont
  {Heisenberg}}, \bibinfo {author} {\bibfnamefont {Ryotaro}\ \bibnamefont
  {Kase}}, \bibinfo {author} {\bibfnamefont {Shinji}\ \bibnamefont
  {Mukohyama}}, \bibinfo {author} {\bibfnamefont {Shinji}\ \bibnamefont
  {Tsujikawa}}, \ and\ \bibinfo {author} {\bibfnamefont {Ying-li}\ \bibnamefont
  {Zhang}},\ }\bibfield  {title} {\enquote {\bibinfo {title} {{Effective
  gravitational couplings for cosmological perturbations in generalized Proca
  theories}},}\ }\href@noop {} {\  (\bibinfo {year} {2016}{\natexlab{c}})},\
  \Eprint {http://arxiv.org/abs/1605.05066} {arXiv:1605.05066 [gr-qc]}
  \BibitemShut {NoStop}%
%%CITATION = ARXIV:1605.05066;%%
\bibitem [{\citenamefont {Horndeski}(1976)}]{Horndeski1976}%
  \BibitemOpen
  \bibfield  {author} {\bibinfo {author} {\bibfnamefont {Gregory~W.}\
  \bibnamefont {Horndeski}},\ }\bibfield  {title} {\enquote {\bibinfo {title}
  {{Conservation of charge and the Einstein-Maxwell field equations}},}\ }\href
  {\doibase 10.1063/1.522837} {\bibfield  {journal} {\bibinfo  {journal} {J.
  Math. Phys.}\ }\textbf {\bibinfo {volume} {17}} (\bibinfo {year} {1976}),\
  10.1063/1.522837}\BibitemShut {NoStop}%
\bibitem [{\citenamefont {Sushkov}(2009)}]{Sushkov:2009hk}%
  \BibitemOpen
  \bibfield  {author} {\bibinfo {author} {\bibfnamefont {Sergey~V.}\
  \bibnamefont {Sushkov}},\ }\bibfield  {title} {\enquote {\bibinfo {title}
  {{Exact cosmological solutions with nonminimal derivative coupling}},}\
  }\href {\doibase 10.1103/PhysRevD.80.103505} {\bibfield  {journal} {\bibinfo
  {journal} {Phys. Rev.}\ }\textbf {\bibinfo {volume} {D80}},\ \bibinfo {pages}
  {103505} (\bibinfo {year} {2009})},\ \Eprint {http://arxiv.org/abs/0910.0980}
  {arXiv:0910.0980 [gr-qc]} \BibitemShut {NoStop}%
%%CITATION = ARXIV:0910.0980;%%
\bibitem [{\citenamefont {Saridakis}\ and\ \citenamefont
  {Sushkov}(2010)}]{Saridakis:2010mf}%
  \BibitemOpen
  \bibfield  {author} {\bibinfo {author} {\bibfnamefont {Emmanuel~N.}\
  \bibnamefont {Saridakis}}\ and\ \bibinfo {author} {\bibfnamefont {Sergey~V.}\
  \bibnamefont {Sushkov}},\ }\bibfield  {title} {\enquote {\bibinfo {title}
  {{Quintessence and phantom cosmology with non-minimal derivative
  coupling}},}\ }\href {\doibase 10.1103/PhysRevD.81.083510} {\bibfield
  {journal} {\bibinfo  {journal} {Phys. Rev.}\ }\textbf {\bibinfo {volume}
  {D81}},\ \bibinfo {pages} {083510} (\bibinfo {year} {2010})},\ \Eprint
  {http://arxiv.org/abs/1002.3478} {arXiv:1002.3478 [gr-qc]} \BibitemShut
  {NoStop}%
%%CITATION = ARXIV:1002.3478;%%
\bibitem [{\citenamefont {Germani}\ and\ \citenamefont
  {Kehagias}(2010)}]{Germani:2010gm}%
  \BibitemOpen
  \bibfield  {author} {\bibinfo {author} {\bibfnamefont {Cristiano}\
  \bibnamefont {Germani}}\ and\ \bibinfo {author} {\bibfnamefont {Alex}\
  \bibnamefont {Kehagias}},\ }\bibfield  {title} {\enquote {\bibinfo {title}
  {{New Model of Inflation with Non-minimal Derivative Coupling of Standard
  Model Higgs Boson to Gravity}},}\ }\href {\doibase
  10.1103/PhysRevLett.105.011302} {\bibfield  {journal} {\bibinfo  {journal}
  {Phys. Rev. Lett.}\ }\textbf {\bibinfo {volume} {105}},\ \bibinfo {pages}
  {011302} (\bibinfo {year} {2010})},\ \Eprint {http://arxiv.org/abs/1003.2635}
  {arXiv:1003.2635 [hep-ph]} \BibitemShut {NoStop}%
%%CITATION = ARXIV:1003.2635;%%
\bibitem [{\citenamefont {Germani}\ and\ \citenamefont
  {Watanabe}(2011)}]{Germani:2011ua}%
  \BibitemOpen
  \bibfield  {author} {\bibinfo {author} {\bibfnamefont {Cristiano}\
  \bibnamefont {Germani}}\ and\ \bibinfo {author} {\bibfnamefont {Yuki}\
  \bibnamefont {Watanabe}},\ }\bibfield  {title} {\enquote {\bibinfo {title}
  {{UV-protected (Natural) Inflation: Primordial Fluctuations and non-Gaussian
  Features}},}\ }\href {\doibase 10.1088/1475-7516/2011/07/031} {\bibfield
  {journal} {\bibinfo  {journal} {JCAP}\ }\textbf {\bibinfo {volume} {1107}},\
  \bibinfo {pages} {031} (\bibinfo {year} {2011})},\ \Eprint
  {http://arxiv.org/abs/1106.0502} {arXiv:1106.0502 [astro-ph.CO]} \BibitemShut
  {NoStop}%
%%CITATION = ARXIV:1106.0502;%%
\bibitem [{\citenamefont {Gubitosi}\ and\ \citenamefont
  {Linder}(2011)}]{Gubitosi:2011sg}%
  \BibitemOpen
  \bibfield  {author} {\bibinfo {author} {\bibfnamefont {Giulia}\ \bibnamefont
  {Gubitosi}}\ and\ \bibinfo {author} {\bibfnamefont {Eric~V.}\ \bibnamefont
  {Linder}},\ }\bibfield  {title} {\enquote {\bibinfo {title} {{Purely Kinetic
  Coupled Gravity}},}\ }\href {\doibase 10.1016/j.physletb.2011.07.066}
  {\bibfield  {journal} {\bibinfo  {journal} {Phys. Lett.}\ }\textbf {\bibinfo
  {volume} {B703}},\ \bibinfo {pages} {113--118} (\bibinfo {year} {2011})},\
  \Eprint {http://arxiv.org/abs/1106.2815} {arXiv:1106.2815 [astro-ph.CO]}
  \BibitemShut {NoStop}%
%%CITATION = ARXIV:1106.2815;%%
\bibitem [{\citenamefont {Babichev}\ and\ \citenamefont
  {Charmousis}(2014)}]{Babichev:2013cya}%
  \BibitemOpen
  \bibfield  {author} {\bibinfo {author} {\bibfnamefont {Eugeny}\ \bibnamefont
  {Babichev}}\ and\ \bibinfo {author} {\bibfnamefont {Christos}\ \bibnamefont
  {Charmousis}},\ }\bibfield  {title} {\enquote {\bibinfo {title} {{Dressing a
  black hole with a time-dependent Galileon}},}\ }\href {\doibase
  10.1007/JHEP08(2014)106} {\bibfield  {journal} {\bibinfo  {journal} {JHEP}\
  }\textbf {\bibinfo {volume} {1408}},\ \bibinfo {pages} {106} (\bibinfo {year}
  {2014})},\ \Eprint {http://arxiv.org/abs/1312.3204} {arXiv:1312.3204 [gr-qc]}
  \BibitemShut {NoStop}%
%%CITATION = ARXIV:1312.3204;%%
\bibitem [{\citenamefont {Rinaldi}(2012)}]{Rinaldi:2012vy}%
  \BibitemOpen
  \bibfield  {author} {\bibinfo {author} {\bibfnamefont {Massimiliano}\
  \bibnamefont {Rinaldi}},\ }\bibfield  {title} {\enquote {\bibinfo {title}
  {{Black holes with non-minimal derivative coupling}},}\ }\href {\doibase
  10.1103/PhysRevD.86.084048} {\bibfield  {journal} {\bibinfo  {journal} {Phys.
  Rev.}\ }\textbf {\bibinfo {volume} {D86}},\ \bibinfo {pages} {084048}
  (\bibinfo {year} {2012})},\ \Eprint {http://arxiv.org/abs/1208.0103}
  {arXiv:1208.0103 [gr-qc]} \BibitemShut {NoStop}%
%%CITATION = ARXIV:1208.0103;%%
\bibitem [{\citenamefont {Minamitsuji}(2014)}]{Minamitsuji:2013ura}%
  \BibitemOpen
  \bibfield  {author} {\bibinfo {author} {\bibfnamefont {Masato}\ \bibnamefont
  {Minamitsuji}},\ }\bibfield  {title} {\enquote {\bibinfo {title} {{Solutions
  in the scalar-tensor theory with nonminimal derivative coupling}},}\ }\href
  {\doibase 10.1103/PhysRevD.89.064017} {\bibfield  {journal} {\bibinfo
  {journal} {Phys. Rev.}\ }\textbf {\bibinfo {volume} {D89}},\ \bibinfo {pages}
  {064017} (\bibinfo {year} {2014})},\ \Eprint {http://arxiv.org/abs/1312.3759}
  {arXiv:1312.3759 [gr-qc]} \BibitemShut {NoStop}%
%%CITATION = ARXIV:1312.3759;%%
\bibitem [{\citenamefont {Anabalon}\ \emph {et~al.}(2014)\citenamefont
  {Anabalon}, \citenamefont {Cisterna},\ and\ \citenamefont
  {Oliva}}]{Anabalon:2013oea}%
  \BibitemOpen
  \bibfield  {author} {\bibinfo {author} {\bibfnamefont {Andres}\ \bibnamefont
  {Anabalon}}, \bibinfo {author} {\bibfnamefont {Adolfo}\ \bibnamefont
  {Cisterna}}, \ and\ \bibinfo {author} {\bibfnamefont {Julio}\ \bibnamefont
  {Oliva}},\ }\bibfield  {title} {\enquote {\bibinfo {title} {{Asymptotically
  locally AdS and flat black holes in Horndeski theory}},}\ }\href {\doibase
  10.1103/PhysRevD.89.084050} {\bibfield  {journal} {\bibinfo  {journal} {Phys.
  Rev.}\ }\textbf {\bibinfo {volume} {D89}},\ \bibinfo {pages} {084050}
  (\bibinfo {year} {2014})},\ \Eprint {http://arxiv.org/abs/1312.3597}
  {arXiv:1312.3597 [gr-qc]} \BibitemShut {NoStop}%
%%CITATION = ARXIV:1312.3597;%%
\bibitem [{\citenamefont {Kobayashi}\ and\ \citenamefont
  {Tanahashi}(2014)}]{Kobayashi:2014eva}%
  \BibitemOpen
  \bibfield  {author} {\bibinfo {author} {\bibfnamefont {Tsutomu}\ \bibnamefont
  {Kobayashi}}\ and\ \bibinfo {author} {\bibfnamefont {Norihiro}\ \bibnamefont
  {Tanahashi}},\ }\bibfield  {title} {\enquote {\bibinfo {title} {{Exact black
  hole solutions in shift symmetric scalar-tensor theories}},}\ }\href
  {\doibase 10.1093/ptep/ptu096} {\bibfield  {journal} {\bibinfo  {journal}
  {PTEP}\ }\textbf {\bibinfo {volume} {2014}},\ \bibinfo {pages} {073E02}
  (\bibinfo {year} {2014})},\ \Eprint {http://arxiv.org/abs/1403.4364}
  {arXiv:1403.4364 [gr-qc]} \BibitemShut {NoStop}%
%%CITATION = ARXIV:1403.4364;%%
\bibitem [{\citenamefont {Charmousis}\ \emph {et~al.}(2014)\citenamefont
  {Charmousis}, \citenamefont {Kolyvaris}, \citenamefont {Papantonopoulos},\
  and\ \citenamefont {Tsoukalas}}]{Charmousis:2014zaa}%
  \BibitemOpen
  \bibfield  {author} {\bibinfo {author} {\bibfnamefont {Christos}\
  \bibnamefont {Charmousis}}, \bibinfo {author} {\bibfnamefont {Theodoros}\
  \bibnamefont {Kolyvaris}}, \bibinfo {author} {\bibfnamefont {Eleftherios}\
  \bibnamefont {Papantonopoulos}}, \ and\ \bibinfo {author} {\bibfnamefont
  {Minas}\ \bibnamefont {Tsoukalas}},\ }\bibfield  {title} {\enquote {\bibinfo
  {title} {{Black Holes in Bi-scalar Extensions of Horndeski Theories}},}\
  }\href {\doibase 10.1007/JHEP07(2014)085} {\bibfield  {journal} {\bibinfo
  {journal} {JHEP}\ }\textbf {\bibinfo {volume} {07}},\ \bibinfo {pages} {085}
  (\bibinfo {year} {2014})},\ \Eprint {http://arxiv.org/abs/1404.1024}
  {arXiv:1404.1024 [gr-qc]} \BibitemShut {NoStop}%
%%CITATION = ARXIV:1404.1024;%%
\bibitem [{\citenamefont {Babichev}\ \emph {et~al.}(2015)\citenamefont
  {Babichev}, \citenamefont {Charmousis},\ and\ \citenamefont
  {Hassaine}}]{Babichev:2015rva}%
  \BibitemOpen
  \bibfield  {author} {\bibinfo {author} {\bibfnamefont {Eugeny}\ \bibnamefont
  {Babichev}}, \bibinfo {author} {\bibfnamefont {Christos}\ \bibnamefont
  {Charmousis}}, \ and\ \bibinfo {author} {\bibfnamefont {Mokhtar}\
  \bibnamefont {Hassaine}},\ }\bibfield  {title} {\enquote {\bibinfo {title}
  {{Charged Galileon black holes}},}\ }\href {\doibase
  10.1088/1475-7516/2015/05/031} {\bibfield  {journal} {\bibinfo  {journal}
  {JCAP}\ }\textbf {\bibinfo {volume} {1505}},\ \bibinfo {pages} {031}
  (\bibinfo {year} {2015})},\ \Eprint {http://arxiv.org/abs/1503.02545}
  {arXiv:1503.02545 [gr-qc]} \BibitemShut {NoStop}%
%%CITATION = ARXIV:1503.02545;%%
\bibitem [{\citenamefont {Silva}\ \emph {et~al.}(2016)\citenamefont {Silva},
  \citenamefont {Maselli}, \citenamefont {Minamitsuji},\ and\ \citenamefont
  {Berti}}]{Silva:2016smx}%
  \BibitemOpen
  \bibfield  {author} {\bibinfo {author} {\bibfnamefont {Hector~O.}\
  \bibnamefont {Silva}}, \bibinfo {author} {\bibfnamefont {Andrea}\
  \bibnamefont {Maselli}}, \bibinfo {author} {\bibfnamefont {Masato}\
  \bibnamefont {Minamitsuji}}, \ and\ \bibinfo {author} {\bibfnamefont
  {Emanuele}\ \bibnamefont {Berti}},\ }\bibfield  {title} {\enquote {\bibinfo
  {title} {{Compact objects in Horndeski gravity}},}\ }in\ \href
  {http://inspirehep.net/record/1422592/files/arXiv:1602.05997.pdf} {\emph
  {\bibinfo {booktitle} {{3rd Amazonian Symposium on Physics and 5th NRHEP
  Network Meeting is approaching: Celebrating 100 Years of General Relativity
  Belem, Brazil, September 28-October 2, 2015}}}}\ (\bibinfo {year} {2016})\
  \Eprint {http://arxiv.org/abs/1602.05997} {arXiv:1602.05997 [gr-qc]}
  \BibitemShut {NoStop}%
%%CITATION = ARXIV:1602.05997;%%
\bibitem [{\citenamefont {Babichev}\ \emph {et~al.}(2016)\citenamefont
  {Babichev}, \citenamefont {Charmousis},\ and\ \citenamefont
  {Lehebel}}]{Babichev:2016rlq}%
  \BibitemOpen
  \bibfield  {author} {\bibinfo {author} {\bibfnamefont {Eugeny}\ \bibnamefont
  {Babichev}}, \bibinfo {author} {\bibfnamefont {Christos}\ \bibnamefont
  {Charmousis}}, \ and\ \bibinfo {author} {\bibfnamefont {Antoine}\
  \bibnamefont {Lehebel}},\ }\bibfield  {title} {\enquote {\bibinfo {title}
  {{Black holes and stars in Horndeski theory}},}\ }\href@noop {} {\  (\bibinfo
  {year} {2016})},\ \Eprint {http://arxiv.org/abs/1604.06402} {arXiv:1604.06402
  [gr-qc]} \BibitemShut {NoStop}%
%%CITATION = ARXIV:1604.06402;%%
\bibitem [{\citenamefont {Bekenstein}(1972{\natexlab{a}})}]{Bekenstein:1971hc}%
  \BibitemOpen
  \bibfield  {author} {\bibinfo {author} {\bibfnamefont {Jacob~D.}\
  \bibnamefont {Bekenstein}},\ }\bibfield  {title} {\enquote {\bibinfo {title}
  {{Nonexistence of baryon number for static black holes}},}\ }\href {\doibase
  10.1103/PhysRevD.5.1239} {\bibfield  {journal} {\bibinfo  {journal} {Phys.
  Rev.}\ }\textbf {\bibinfo {volume} {D5}},\ \bibinfo {pages} {1239--1246}
  (\bibinfo {year} {1972}{\natexlab{a}})}\BibitemShut {NoStop}%
%%CITATION = PHRVA,D5,1239;%%
\bibitem [{\citenamefont {Bekenstein}(1972{\natexlab{b}})}]{Bekenstein:1972ky}%
  \BibitemOpen
  \bibfield  {author} {\bibinfo {author} {\bibfnamefont {J.D.}\ \bibnamefont
  {Bekenstein}},\ }\bibfield  {title} {\enquote {\bibinfo {title}
  {{Nonexistence of baryon number for black holes. ii}},}\ }\href {\doibase
  10.1103/PhysRevD.5.2403} {\bibfield  {journal} {\bibinfo  {journal}
  {Phys.Rev.}\ }\textbf {\bibinfo {volume} {D5}},\ \bibinfo {pages}
  {2403--2412} (\bibinfo {year} {1972}{\natexlab{b}})}\BibitemShut {NoStop}%
%%CITATION = PHRVA,D5,2403;%%
\bibitem [{\citenamefont {Chagoya}\ \emph {et~al.}(2016)\citenamefont
  {Chagoya}, \citenamefont {Niz},\ and\ \citenamefont
  {Tasinato}}]{Chagoya:2016aar}%
  \BibitemOpen
  \bibfield  {author} {\bibinfo {author} {\bibfnamefont {Javier}\ \bibnamefont
  {Chagoya}}, \bibinfo {author} {\bibfnamefont {Gustavo}\ \bibnamefont {Niz}},
  \ and\ \bibinfo {author} {\bibfnamefont {Gianmassimo}\ \bibnamefont
  {Tasinato}},\ }\bibfield  {title} {\enquote {\bibinfo {title} {{Black Holes
  and Abelian Symmetry Breaking}},}\ }\href@noop {} {\  (\bibinfo {year}
  {2016})},\ \Eprint {http://arxiv.org/abs/1602.08697} {arXiv:1602.08697
  [hep-th]} \BibitemShut {NoStop}%
%%CITATION = ARXIV:1602.08697;%%
\bibitem [{\citenamefont {Herdeiro}\ \emph {et~al.}(2016)\citenamefont
  {Herdeiro}, \citenamefont {Radu},\ and\ \citenamefont
  {Runarsson}}]{Herdeiro:2016tmi}%
  \BibitemOpen
  \bibfield  {author} {\bibinfo {author} {\bibfnamefont {Carlos}\ \bibnamefont
  {Herdeiro}}, \bibinfo {author} {\bibfnamefont {Eugen}\ \bibnamefont {Radu}},
  \ and\ \bibinfo {author} {\bibfnamefont {Helgi}\ \bibnamefont {Runarsson}},\
  }\bibfield  {title} {\enquote {\bibinfo {title} {{Kerr black holes with Proca
  hair}},}\ }\href {\doibase 10.1088/0264-9381/33/15/154001} {\bibfield
  {journal} {\bibinfo  {journal} {Class. Quant. Grav.}\ }\textbf {\bibinfo
  {volume} {33}},\ \bibinfo {pages} {154001} (\bibinfo {year} {2016})},\
  \Eprint {http://arxiv.org/abs/1603.02687} {arXiv:1603.02687 [gr-qc]}
  \BibitemShut {NoStop}%
%%CITATION = ARXIV:1603.02687;%%
\bibitem [{\citenamefont {Geng}\ and\ \citenamefont {Lu}(2016)}]{Geng:2015kvs}%
  \BibitemOpen
  \bibfield  {author} {\bibinfo {author} {\bibfnamefont {Wei-Jian}\
  \bibnamefont {Geng}}\ and\ \bibinfo {author} {\bibfnamefont {H.}~\bibnamefont
  {Lu}},\ }\bibfield  {title} {\enquote {\bibinfo {title} {{Einstein-Vector
  Gravity, Emerging Gauge Symmetry and de Sitter Bounce}},}\ }\href {\doibase
  10.1103/PhysRevD.93.044035} {\bibfield  {journal} {\bibinfo  {journal} {Phys.
  Rev.}\ }\textbf {\bibinfo {volume} {D93}},\ \bibinfo {pages} {044035}
  (\bibinfo {year} {2016})},\ \Eprint {http://arxiv.org/abs/1511.03681}
  {arXiv:1511.03681 [hep-th]} \BibitemShut {NoStop}%
%%CITATION = ARXIV:1511.03681;%%
\bibitem [{\citenamefont {Beltran~Jimenez}\ and\ \citenamefont
  {Koivisto}(2014)}]{Jimenez:2014rna}%
  \BibitemOpen
  \bibfield  {author} {\bibinfo {author} {\bibfnamefont {Jose}\ \bibnamefont
  {Beltran~Jimenez}}\ and\ \bibinfo {author} {\bibfnamefont {Tomi~S.}\
  \bibnamefont {Koivisto}},\ }\bibfield  {title} {\enquote {\bibinfo {title}
  {{Extended Gauss-Bonnet gravities in Weyl geometry}},}\ }\href {\doibase
  10.1088/0264-9381/31/13/135002} {\bibfield  {journal} {\bibinfo  {journal}
  {Class. Quant. Grav.}\ }\textbf {\bibinfo {volume} {31}},\ \bibinfo {pages}
  {135002} (\bibinfo {year} {2014})},\ \Eprint {http://arxiv.org/abs/1402.1846}
  {arXiv:1402.1846 [gr-qc]} \BibitemShut {NoStop}%
%%CITATION = ARXIV:1402.1846;%%
\bibitem [{\citenamefont {Fan}(2016)}]{Fan:2016jnz}%
  \BibitemOpen
  \bibfield  {author} {\bibinfo {author} {\bibfnamefont {Zhong-Ying}\
  \bibnamefont {Fan}},\ }\bibfield  {title} {\enquote {\bibinfo {title} {{Black
  Holes With Vector Hair}},}\ }\href@noop {} {\  (\bibinfo {year} {2016})},\
  \Eprint {http://arxiv.org/abs/1606.00684} {arXiv:1606.00684 [hep-th]}
  \BibitemShut {NoStop}%
%%CITATION = ARXIV:1606.00684;%%
\bibitem [{\citenamefont {Hartle}(1967)}]{Hartle:1967he}%
  \BibitemOpen
  \bibfield  {author} {\bibinfo {author} {\bibfnamefont {James~B.}\
  \bibnamefont {Hartle}},\ }\bibfield  {title} {\enquote {\bibinfo {title}
  {{Slowly rotating relativistic stars. I. Equations of structure}},}\ }\href
  {\doibase 10.1086/149400} {\bibfield  {journal} {\bibinfo  {journal}
  {Astrophys.J.}\ }\textbf {\bibinfo {volume} {150}},\ \bibinfo {pages}
  {1005--1029} (\bibinfo {year} {1967})}\BibitemShut {NoStop}%
%%CITATION = ASJOA,150,1005;%%
\bibitem [{\citenamefont {Hartle}\ and\ \citenamefont
  {Thorne}(1968)}]{Hartle:1968si}%
  \BibitemOpen
  \bibfield  {author} {\bibinfo {author} {\bibfnamefont {James~B.}\
  \bibnamefont {Hartle}}\ and\ \bibinfo {author} {\bibfnamefont {Kip~S.}\
  \bibnamefont {Thorne}},\ }\bibfield  {title} {\enquote {\bibinfo {title}
  {{Slowly Rotating Relativistic Stars. II. Models for Neutron Stars and
  Supermassive Stars}},}\ }\href {\doibase 10.1086/149707} {\bibfield
  {journal} {\bibinfo  {journal} {Astrophys.J.}\ }\textbf {\bibinfo {volume}
  {153}},\ \bibinfo {pages} {807} (\bibinfo {year} {1968})}\BibitemShut
  {NoStop}%
%%CITATION = ASJOA,153,807;%%
\bibitem [{\citenamefont {Maselli}\ \emph {et~al.}(2015)\citenamefont
  {Maselli}, \citenamefont {Silva}, \citenamefont {Minamitsuji},\ and\
  \citenamefont {Berti}}]{Maselli:2015yva}%
  \BibitemOpen
  \bibfield  {author} {\bibinfo {author} {\bibfnamefont {Andrea}\ \bibnamefont
  {Maselli}}, \bibinfo {author} {\bibfnamefont {Hector~O.}\ \bibnamefont
  {Silva}}, \bibinfo {author} {\bibfnamefont {Masato}\ \bibnamefont
  {Minamitsuji}}, \ and\ \bibinfo {author} {\bibfnamefont {Emanuele}\
  \bibnamefont {Berti}},\ }\bibfield  {title} {\enquote {\bibinfo {title}
  {{Slowly rotating black hole solutions in Horndeski gravity}},}\ }\href
  {\doibase 10.1103/PhysRevD.92.104049} {\bibfield  {journal} {\bibinfo
  {journal} {Phys. Rev.}\ }\textbf {\bibinfo {volume} {D92}},\ \bibinfo {pages}
  {104049} (\bibinfo {year} {2015})},\ \Eprint
  {http://arxiv.org/abs/1508.03044} {arXiv:1508.03044 [gr-qc]} \BibitemShut
  {NoStop}%
%%CITATION = ARXIV:1508.03044;%%
\bibitem [{\citenamefont {Cisterna}\ \emph
  {et~al.}(2015{\natexlab{a}})\citenamefont {Cisterna}, \citenamefont {Cruz},
  \citenamefont {Delsate},\ and\ \citenamefont {Saavedra}}]{Cisterna:2015uya}%
  \BibitemOpen
  \bibfield  {author} {\bibinfo {author} {\bibfnamefont {Adolfo}\ \bibnamefont
  {Cisterna}}, \bibinfo {author} {\bibfnamefont {Miguel}\ \bibnamefont {Cruz}},
  \bibinfo {author} {\bibfnamefont {T\'rence}\ \bibnamefont {Delsate}}, \ and\
  \bibinfo {author} {\bibfnamefont {Joel}\ \bibnamefont {Saavedra}},\
  }\bibfield  {title} {\enquote {\bibinfo {title} {{Nonminimal derivative
  coupling scalar-tensor theories: odd-parity perturbations and black hole
  stability}},}\ }\href {\doibase 10.1103/PhysRevD.92.104018} {\bibfield
  {journal} {\bibinfo  {journal} {Phys. Rev.}\ }\textbf {\bibinfo {volume}
  {D92}},\ \bibinfo {pages} {104018} (\bibinfo {year} {2015}{\natexlab{a}})},\
  \Eprint {http://arxiv.org/abs/1508.06413} {arXiv:1508.06413 [gr-qc]}
  \BibitemShut {NoStop}%
%%CITATION = ARXIV:1508.06413;%%
\bibitem [{\citenamefont {Dehghani}\ and\ \citenamefont
  {KhajehAzad}(2003)}]{Dehghani:2002nt}%
  \BibitemOpen
  \bibfield  {author} {\bibinfo {author} {\bibfnamefont {M.~H.}\ \bibnamefont
  {Dehghani}}\ and\ \bibinfo {author} {\bibfnamefont {H.}~\bibnamefont
  {KhajehAzad}},\ }\bibfield  {title} {\enquote {\bibinfo {title}
  {{Thermodynamics of Kerr-Newman de Sitter black hole and dS / CFT
  correspondence}},}\ }\href {\doibase 10.1139/p03-110} {\bibfield  {journal}
  {\bibinfo  {journal} {Can. J. Phys.}\ }\textbf {\bibinfo {volume} {81}},\
  \bibinfo {pages} {1363} (\bibinfo {year} {2003})},\ \Eprint
  {http://arxiv.org/abs/hep-th/0209203} {arXiv:hep-th/0209203 [hep-th]}
  \BibitemShut {NoStop}%
%%CITATION = HEP-TH/0209203;%%
\bibitem [{\citenamefont {Huaifan}\ \emph {et~al.}(2009)\citenamefont
  {Huaifan}, \citenamefont {Shengli}, \citenamefont {Yueqin}, \citenamefont
  {Lichun},\ and\ \citenamefont {Ren}}]{Huaifan:2009nf}%
  \BibitemOpen
  \bibfield  {author} {\bibinfo {author} {\bibfnamefont {Li}~\bibnamefont
  {Huaifan}}, \bibinfo {author} {\bibfnamefont {Zhang}\ \bibnamefont
  {Shengli}}, \bibinfo {author} {\bibfnamefont {Wu}~\bibnamefont {Yueqin}},
  \bibinfo {author} {\bibfnamefont {Zhang}\ \bibnamefont {Lichun}}, \ and\
  \bibinfo {author} {\bibfnamefont {Zhao}\ \bibnamefont {Ren}},\ }\bibfield
  {title} {\enquote {\bibinfo {title} {{Hawking radiation of Kerr-Newman-de
  Sitter black hole}},}\ }\href {\doibase 10.1140/epjc/s10052-009-1085-0}
  {\bibfield  {journal} {\bibinfo  {journal} {Eur. Phys. J.}\ }\textbf
  {\bibinfo {volume} {C63}},\ \bibinfo {pages} {133--138} (\bibinfo {year}
  {2009})},\ \Eprint {http://arxiv.org/abs/0906.3680} {arXiv:0906.3680 [gr-qc]}
  \BibitemShut {NoStop}%
%%CITATION = ARXIV:0906.3680;%%
\bibitem [{\citenamefont {Kobayashi}\ \emph {et~al.}(2012)\citenamefont
  {Kobayashi}, \citenamefont {Motohashi},\ and\ \citenamefont
  {Suyama}}]{Kobayashi:2012kh}%
  \BibitemOpen
  \bibfield  {author} {\bibinfo {author} {\bibfnamefont {Tsutomu}\ \bibnamefont
  {Kobayashi}}, \bibinfo {author} {\bibfnamefont {Hayato}\ \bibnamefont
  {Motohashi}}, \ and\ \bibinfo {author} {\bibfnamefont {Teruaki}\ \bibnamefont
  {Suyama}},\ }\bibfield  {title} {\enquote {\bibinfo {title} {{Black hole
  perturbation in the most general scalar-tensor theory with second-order field
  equations I: the odd-parity sector}},}\ }\href {\doibase
  10.1103/PhysRevD.85.084025} {\bibfield  {journal} {\bibinfo  {journal} {Phys.
  Rev.}\ }\textbf {\bibinfo {volume} {D85}},\ \bibinfo {pages} {084025}
  (\bibinfo {year} {2012})},\ \Eprint {http://arxiv.org/abs/1202.4893}
  {arXiv:1202.4893 [gr-qc]} \BibitemShut {NoStop}%
%%CITATION = ARXIV:1202.4893;%%
\bibitem [{\citenamefont {Kobayashi}\ \emph {et~al.}(2014)\citenamefont
  {Kobayashi}, \citenamefont {Motohashi},\ and\ \citenamefont
  {Suyama}}]{Kobayashi:2014wsa}%
  \BibitemOpen
  \bibfield  {author} {\bibinfo {author} {\bibfnamefont {Tsutomu}\ \bibnamefont
  {Kobayashi}}, \bibinfo {author} {\bibfnamefont {Hayato}\ \bibnamefont
  {Motohashi}}, \ and\ \bibinfo {author} {\bibfnamefont {Teruaki}\ \bibnamefont
  {Suyama}},\ }\bibfield  {title} {\enquote {\bibinfo {title} {{Black hole
  perturbation in the most general scalar-tensor theory with second-order field
  equations. II. The even-parity sector}},}\ }\href {\doibase
  10.1103/PhysRevD.89.084042} {\bibfield  {journal} {\bibinfo  {journal} {Phys.
  Rev.}\ }\textbf {\bibinfo {volume} {D89}},\ \bibinfo {pages} {084042}
  (\bibinfo {year} {2014})},\ \Eprint {http://arxiv.org/abs/1402.6740}
  {arXiv:1402.6740 [gr-qc]} \BibitemShut {NoStop}%
%%CITATION = ARXIV:1402.6740;%%
\bibitem [{\citenamefont {Ogawa}\ \emph {et~al.}(2016)\citenamefont {Ogawa},
  \citenamefont {Kobayashi},\ and\ \citenamefont {Suyama}}]{Ogawa:2015pea}%
  \BibitemOpen
  \bibfield  {author} {\bibinfo {author} {\bibfnamefont {Hiromu}\ \bibnamefont
  {Ogawa}}, \bibinfo {author} {\bibfnamefont {Tsutomu}\ \bibnamefont
  {Kobayashi}}, \ and\ \bibinfo {author} {\bibfnamefont {Teruaki}\ \bibnamefont
  {Suyama}},\ }\bibfield  {title} {\enquote {\bibinfo {title} {{Instability of
  hairy black holes in shift-symmetric Horndeski theories}},}\ }\href {\doibase
  10.1103/PhysRevD.93.064078} {\bibfield  {journal} {\bibinfo  {journal} {Phys.
  Rev.}\ }\textbf {\bibinfo {volume} {D93}},\ \bibinfo {pages} {064078}
  (\bibinfo {year} {2016})},\ \Eprint {http://arxiv.org/abs/1510.07400}
  {arXiv:1510.07400 [gr-qc]} \BibitemShut {NoStop}%
%%CITATION = ARXIV:1510.07400;%%
\bibitem [{\citenamefont {Cisterna}\ \emph
  {et~al.}(2015{\natexlab{b}})\citenamefont {Cisterna}, \citenamefont
  {Delsate},\ and\ \citenamefont {Rinaldi}}]{Cisterna:2015yla}%
  \BibitemOpen
  \bibfield  {author} {\bibinfo {author} {\bibfnamefont {Adolfo}\ \bibnamefont
  {Cisterna}}, \bibinfo {author} {\bibfnamefont {T{\'e}rence}\ \bibnamefont
  {Delsate}}, \ and\ \bibinfo {author} {\bibfnamefont {Massimiliano}\
  \bibnamefont {Rinaldi}},\ }\bibfield  {title} {\enquote {\bibinfo {title}
  {{Neutron stars in general second order scalar-tensor theory: The case of
  nonminimal derivative coupling}},}\ }\href {\doibase
  10.1103/PhysRevD.92.044050} {\bibfield  {journal} {\bibinfo  {journal} {Phys.
  Rev.}\ }\textbf {\bibinfo {volume} {D92}},\ \bibinfo {pages} {044050}
  (\bibinfo {year} {2015}{\natexlab{b}})},\ \Eprint
  {http://arxiv.org/abs/1504.05189} {arXiv:1504.05189 [gr-qc]} \BibitemShut
  {NoStop}%
%%CITATION = ARXIV:1504.05189;%%
\bibitem [{\citenamefont {Cisterna}\ \emph {et~al.}(2016)\citenamefont
  {Cisterna}, \citenamefont {Delsate}, \citenamefont {Ducobu},\ and\
  \citenamefont {Rinaldi}}]{Cisterna:2016vdx}%
  \BibitemOpen
  \bibfield  {author} {\bibinfo {author} {\bibfnamefont {Adolfo}\ \bibnamefont
  {Cisterna}}, \bibinfo {author} {\bibfnamefont {Terence}\ \bibnamefont
  {Delsate}}, \bibinfo {author} {\bibfnamefont {Ludovic}\ \bibnamefont
  {Ducobu}}, \ and\ \bibinfo {author} {\bibfnamefont {Massimiliano}\
  \bibnamefont {Rinaldi}},\ }\bibfield  {title} {\enquote {\bibinfo {title}
  {{Slowly rotating neutron stars in the nonminimal derivative coupling sector
  of Horndeski gravity}},}\ }\href {\doibase 10.1103/PhysRevD.93.084046}
  {\bibfield  {journal} {\bibinfo  {journal} {Phys. Rev.}\ }\textbf {\bibinfo
  {volume} {D93}},\ \bibinfo {pages} {084046} (\bibinfo {year} {2016})},\
  \Eprint {http://arxiv.org/abs/1602.06939} {arXiv:1602.06939 [gr-qc]}
  \BibitemShut {NoStop}%
%%CITATION = ARXIV:1602.06939;%%
\bibitem [{\citenamefont {Maselli}\ \emph {et~al.}(2016)\citenamefont
  {Maselli}, \citenamefont {Silva}, \citenamefont {Minamitsuji},\ and\
  \citenamefont {Berti}}]{Maselli:2016gxk}%
  \BibitemOpen
  \bibfield  {author} {\bibinfo {author} {\bibfnamefont {Andrea}\ \bibnamefont
  {Maselli}}, \bibinfo {author} {\bibfnamefont {Hector~O.}\ \bibnamefont
  {Silva}}, \bibinfo {author} {\bibfnamefont {Masato}\ \bibnamefont
  {Minamitsuji}}, \ and\ \bibinfo {author} {\bibfnamefont {Emanuele}\
  \bibnamefont {Berti}},\ }\bibfield  {title} {\enquote {\bibinfo {title}
  {{Neutron stars in Horndeski gravity}},}\ }\href {\doibase
  10.1103/PhysRevD.93.124056} {\bibfield  {journal} {\bibinfo  {journal} {Phys.
  Rev.}\ }\textbf {\bibinfo {volume} {D93}},\ \bibinfo {pages} {124056}
  (\bibinfo {year} {2016})},\ \Eprint {http://arxiv.org/abs/1603.04876}
  {arXiv:1603.04876 [gr-qc]} \BibitemShut {NoStop}%
%%CITATION = ARXIV:1603.04876;%%
\bibitem [{\citenamefont {Brihaye}\ \emph {et~al.}(2016)\citenamefont
  {Brihaye}, \citenamefont {Cisterna},\ and\ \citenamefont
  {Erices}}]{Brihaye:2016lin}%
  \BibitemOpen
  \bibfield  {author} {\bibinfo {author} {\bibfnamefont {Yves}\ \bibnamefont
  {Brihaye}}, \bibinfo {author} {\bibfnamefont {Adolfo}\ \bibnamefont
  {Cisterna}}, \ and\ \bibinfo {author} {\bibfnamefont {Cristian}\ \bibnamefont
  {Erices}},\ }\bibfield  {title} {\enquote {\bibinfo {title} {{Boson stars in
  biscalar extensions of Horndeski gravity}},}\ }\href {\doibase
  10.1103/PhysRevD.93.124057} {\bibfield  {journal} {\bibinfo  {journal} {Phys.
  Rev.}\ }\textbf {\bibinfo {volume} {D93}},\ \bibinfo {pages} {124057}
  (\bibinfo {year} {2016})},\ \Eprint {http://arxiv.org/abs/1604.02121}
  {arXiv:1604.02121 [hep-th]} \BibitemShut {NoStop}%
%%CITATION = ARXIV:1604.02121;%%
\end{thebibliography}%
\end{document}